\documentclass[iop]{emulateapj}
\usepackage{graphicx}

\usepackage{natbib}
\shorttitle{The Stability of LSB Disks}
\shortauthors{MacLachlan et al.}
\newcommand{\HI}{\mbox{H\,{\sc i}}}
\newcommand{\HII}{\mbox{H\,{\sc i}{\sc i}}}
\newcommand{\gsim}{\mathrel{\hbox{\rlap{\lower.55ex 
\hbox{$\sim$}} \kern-.3em \raise.4ex \hbox{$>$}}}}
\newcommand{\lsim}{\mathrel{\hbox{\rlap{\lower.55ex 
\hbox{$\sim$}} \kern-.3em \raise.4ex \hbox{$<$}}}}
\begin{document}
\title{The Stability of Low Surface Brightness Disks Based on Multi-Wavelength Modeling}

\author{J. M. MacLachlan}
\affil{School of Physics and Astronomy, University of St Andrews, North Haugh,
St Andrews, Fife, KY16 9SS, Scotland}
\email{jmm55@st-andrews.ac.uk}

\author{L. D. Matthews\altaffilmark{1}}
\affil{Harvard-Smithsonian Center for Astrophysics, 60 Garden Street, MS-42, Cambridge, MA 02138, USA}

\author{K. Wood}
\affil{School of Physics and Astronomy, University of St Andrews, North Haugh,
St Andrews, Fife, KY16 9SS, Scotland}
\and
\author{J. S. Gallagher}
\affil{Department of Astronomy, University of Wisconsin, 475 N. Charter Street, Madison, WI, 53706, USA}

\altaffiltext{1}{Current address: Massachusetts Institute of Technology, Haystack Observatory, Off Route 40, Westford, MA 01886, USA}
\begin{abstract}

To investigate the structure and composition of the dusty interstellar medium (ISM) of low surface brightness (LSB) disk galaxies, we have used multiwavelength photometry to construct spectral energy distributions for three low-mass, edge-on LSB galaxies ($V_{\rm rot}$= 88 to 105 km s$^{-1}$).  We use Monte Carlo radiation transfer codes that include the effects of transiently heated small grains and polycyclic aromatic hydrocarbon molecules to model and interpret the data. We find that unlike the high surface brightness galaxies previously modeled, the dust disks appear to have scale heights equal to or exceeding their stellar scale heights. This result supports the findings of previous studies that low mass disk galaxies have dust scale heights comparable to their stellar scale heights and suggests that the cold ISM of low mass, LSB disk galaxies may be stable against fragmentation and gravitational collapse. This may help to explain the lack of observed dust lanes in edge-on LSB galaxies and their low current star formation rates. Dust masses are found in the range $1.16$ to $2.38\times10^{6}M_{\odot}$, corresponding to face-on (edge-on), $V$-band, optical depths $0.034 \lsim \tau_{face} \lsim 0.106$ ($0.69 \lsim \tau_{eq} \lsim 1.99$).
\end{abstract}

\section{Introduction}
Low surface brightness (LSB) disk galaxies, whose central, face-on surface brightnesses fall below $\mu_{B,0}\gsim$ 23.0 mag arcsec$^{-2}$ in the $B$ band, account for an important fraction of the luminosity and galactic mass densities of the local universe \citep{driver_contribution_1999,minchin_cosmological_2004,oneil_space_2000}. LSB disk galaxies show a range of masses and morphologies \citep[e.g., ][]{sprayberry_low_1997,auld_morphology_2006}, but the most common are bulgeless, late-type LSB disks with Hubble types Sd-Sdm. These galaxies frequently show signs of being dark matter dominated at nearly all radii \citep[e.g., ][]{de_blok_dark_1997,de_blok_high-resolution_2002,banerjee_dark_2010}.

LSB galaxies have been extensively studied in the \HI\ 21-cm line. Often their total \HI\ masses are comparable to or larger than their high surface brightness (HSB) counterparts, while their \HI\ surface densities are lower by a factor of $\sim$2 \citep{van_der_hulst_star_1993}. In fact, LSB galaxies are usually found to have most of their \HI{} disk at surface densities below the \citet{kennicutt_star_1989} critical values for the formation of massive stars \citep{van_der_hulst_star_1993}. This fact may lead to a lower than expected star formation efficiency causing LSB galaxies to evolve more slowly than HSB galaxies. A lack of high mass stars will slow metal production and indeed \citet{mcgaugh_oxygen_1994} find LSB galaxies have metal abundances of $\sim1/3$ solar. \HII{} regions are observed in LSB galaxy disks as traced by H$\alpha$ emission \citep{rand_diffuse_1996,mcgaugh_morphology_1995,matthews_extraordinary_1999} and blue disk colours indicating young stars are present  \citep{mcgaugh_morphology_1995,matthews_extraordinary_1999,matthews_H_2008} suggesting LSB galaxies do have ongoing star formation. However, the conditions and structure of the interstellar medium (ISM) in LSB disk galaxies are still uncertain.

While observations of H$\alpha$ and \HI{} have revealed information on the ionized and atomic gas, little is known about the dust and molecular hydrogen in the ISM. Based on observations of $^{12}$CO(1-0) emission in several LSB disk galaxies \citet{matthews_detections_2005} find that the CO emission, and hence the molecular hydrogen content, depends strongly on the rotational velocity. It appears that the rotational velocity, as well as the surface density, play an important role in shaping the structure and content of the ISM. This is also suggested by the work of \citet{dalcanton_formation_2004} who find that the gas and dust disk of a galaxy is more stable to the effects of radial instabilities at lower rotational velocity. This limits the ability of perturbations to cause the collapse of the ISM into a thinner layer and may inhibit global star formation by preventing the formation of large numbers of giant molecular clouds. 

Due to the difficulties of directly detecting molecular gas in low-mass LSB spirals \citep[cf.][]{matthews_co_2001,matthews_detections_2005,das_CO_2006}, an alternative method to investigate the composition and structure of their molecular ISM is to observe far-infrared (FIR) emission produced by dust. The dust and gas are expected to be well mixed and so the highest dust densities should trace the high density, cool phases of the ISM. In order to interpret the distribution of FIR emission we require a radiation transfer model to produce synthetic images and spectral energy distributions based on galactic stellar and dust distributions. Previous modeling efforts have focussed on fitting the optical and near-infrared (NIR) extinction of HSB galaxies \citep{kylafis_dust_1987,xilouris_distribution_1997,xilouris_optical_1998,xilouris_are_1999, bianchi_effects_2000,bianchi_dust_2007} and treating the dust absorption and reemission processes to model the full UV/optical to sub-mm spectral energy distribution (SED) \citep{popescu_modelling_2000,misiriotis_modeling_2001,bianchi_dust_2008,baes_herschel-atlas:_2010,popescu_modelling_2011}. In the present work, we extend this type of analysis for the first time to a sample of LSB galaxies.

When modeling the optical appearance of a disk galaxy an edge-on orientation is advantageous. This viewing geometry allows both the vertical and radial ISM structure to be constrained simultaneously and small scale features of the galaxy are smoothed out due to overlapping sight lines. The effects of dust become more apparent and dark dust lanes can often be seen obscuring stellar emission where the central regions of the galactic disk have become optically thick. This makes the quantification of dust masses and distributions much more straightforward than the more degenerate case of a less inclined disk that is optically thin across most of its area. 

In this paper we further constrain the effects of dust, ISM geometry and mass on the UV/optical - FIR SEDs of edge-on LSB galaxies using Monte Carlo radiation transfer techniques. In section \ref{galsam} we will introduce the sample of galaxies to which the analysis is applied followed by a description of the the data available at various wavelengths in section \ref{data}. Section \ref{code} includes a description of the model used in the analysis. The results for our sample of galaxies can be found in section \ref{results} and a discussion of the key results is presented in section \ref{discuss}.

\section{Galaxy Sample}\label{galsam}
We have selected three LSB galaxies that have been observed at a wide range of wavelengths to allow the best determination of the shape of the SED. The galaxies are nearby ($\le$10~Mpc) and are well-resolved in optical imaging . We have selected the galaxies to span a range of central face-on surface brightness ($\mu_{B,0}=22.6 - 23.6$ mag arcsec$^{-2}$) as well as rotational velocity ($V_{\rm rot}=88 - 105 $ km~$s^{-1}$). The three galaxies have similar total \HI{} masses but observations indicate that their molecular hydrogen content decreases with decreasing rotational velocity ($V_{\rm rot}$) \citep{matthews_detections_2005}, suggesting that the ISM conditions may be different across our sample. Table \ref{prop} contains the properties of the galaxies as gathered from the literature. High resolution optical imaging of the three galaxies can be seen in Figure \ref{HST}.

\subsection{UGC 7321}
UGC 7321 is one of the best studied edge-on LSB galaxies. It shows a very thin disk ($a/b \sim 10.3)$\footnote{a/b is the disk axial ratio measured at the $25.0$ mag arcsec$^{-1}$ $R$-band isophote} with signs of clumping of the stellar and dust distributions in the inner galaxy \citep{matthews_modeling_2001} and a more uniform structure in the outer disk. The disk shows strong ($B-R$) colour gradients in both horizontal and vertical directions that cannot be explained by the effects of dust \citep{matthews_modeling_2001} and appear to be due to stellar population or metallicity gradients within the disk itself. H$\alpha$ imaging shows ionized gas is present in a very thin layer that has very little structure and shows no signs of large scale star forming regions \citep{matthews_extraordinary_1999}. The \HI{} gas content of UGC 7321 has a greater radial extent than the optical disk of the galaxy \citep{uson_H_2003} and shows indications that the \HI{} disk is also vertically extended \citep{matthews_high-latitude_2003}.  Observations of the CO(1-0) line emission by \citet{matthews_co_2001} confirm the presence of molecular gas in UGC 7321 and this measurement leads to a molecular hydrogen mass of $ M_{H_{2}} \sim 3.2\times10^{7}M_{\odot}$ assuming a Galactic CO-H$_{2}$ conversion factor $X = 3.0\times10^{-20}$cm$^{-2}$ (K kms$^{-1}$)$^{-1}$\citep{young_molecular_1991}.
\subsection{IC 2233} 
IC 2233 is another example of a thin, bulge-less galaxy viewed close to edge on. It shows no signs of a dust lane in the central regions of its disk with only a few dark clouds visible. \citet{rand_diffuse_1996} detected H$\alpha$ emission associated with \HII{} regions across the disk of IC 2233, including a large complex in the outer region of the galaxy, indicating that there is ongoing star formation. In contrast to UGC 7321 the \HII{} regions in IC 2233 are often located out of the mid-plane at up to $\sim 0.5Z_{s}$ in the outer parts of the galaxy \citep{matthews_H_2008}. Observations by \citet{matthews_H_2008} find that \HI{} in IC 2233 is extended vertically and horizontally beyond the optical disk, with evidence for flaring in the outer gas disk. Attempts by \citet{matthews_detections_2005} to detect CO emission from IC 2233 were unsuccessful leading to a 3-$\sigma$ upper limit on the mass of molecular hydrogen of $1.4\times 10^{6} M_{\odot}$.
\subsection{NGC 4244}
NGC 4244 is the closest galaxy in our sample lying at an estimated distance of only $\sim 4.4 $Mpc  \citep{seth_study_2005-1}. NGC 4244 shows many dark clouds in optical imaging of its central regions. Despite this, based on NIR photometry \citet{kodaira_near-infrared_1996} conclude that the disk of NGC 4244 suffers from little internal extinction. Continuum subtracted H$\alpha$ imaging obtained by \citet{hoopes_diffuse_1999} shows \HII{} regions extending up to $\sim990$pc from the disk midplane\footnote{This value has been re-scaled to the distance adopted in this paper ($4.4$Mpc)}. The two largest \HII{} regions are found in the outer galactic disk. The work of \citet{olling_ngc_1996} investigating the \HI{} distribution in NGC 4244 shows that the gas distribution appears to be slightly extended radially compared to the stellar disk. Both \citet{matthews_co_2001} and \citet{sage_molecular_1993} report the detection of CO emission from NGC 4244 and the estimated H$_{2}$ mass is $1.4\times10^{7}$M$_{\odot}$.
\begin{figure}
\leavevmode
\includegraphics[width=240pt]{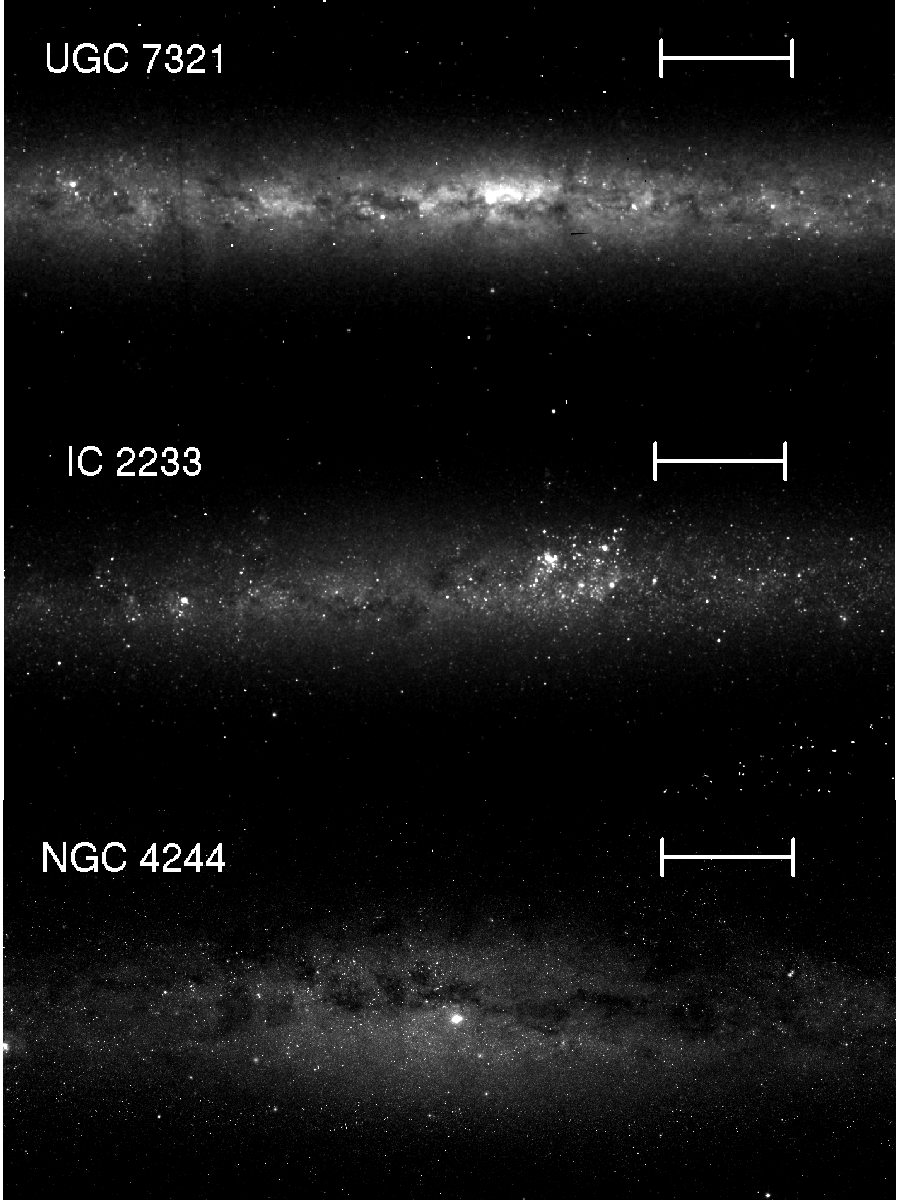}
\caption{High resolution imaging of the inner regions (approximately $\sim 3.4 \times 1.5$kpc)  of the three galaxies studied. WFPC2 $F702W$ and $F814W$ composite image of UGC 7321 ({\it top}). ACS WFC $F606W$ filter images of IC 2233 ({\it middle}) and NGC 4244 ({\it bottom}). The white bar in the top right corner of each image has a length of $500$pc, assuming the distances in Table \ref{prop}. UGC 7321 and NGC 4244 show the presence of dark clouds in their nuclear regions that are seen in lower numbers in IC 2233. This may be indicative of a larger mass of molecular gas in these galaxies which could have an impact on the FIR emission if dust is also associated with these clouds.}
\label{HST}
\end{figure}
\section{Data}\label{data}
All three galaxies have been observed in the optical by the Sloan Digital Sky Survey (SDSS) in $u,g,r,i$ and $z$ bands and the data are available under data release seven\footnote{http://www.SDSS.org/dr7/} \citep{abazajian_seventh_2009}. The images were downloaded from the SDSS archive and fluxes in each band were extracted using GAIA's aperture photometry tool. It was necessary to re-analyse the SDSS images to extract reliable photometry as each galaxy appeared to have been split into multiple components during the SDSS photometric reduction.

We have also made use of $B$ band imaging of UGC 7321 and IC 2233 obtained at the 3.5m WIYN telescope at Kitt Peak, AZ. These observations are described in \citet{matthews_extraordinary_1999} (UGC 7321) and \citet{matthews_H_2008} (IC2233). The images have been flux calibrated and are of a higher resolution than the SDSS imaging available. In the case of NGC 4244 we do not have access to higher resolution data and so the SDSS $r$ band imaging will be used for the optical comparisons.  

High resolution {\it Hubble Space Telescope} ({\it HST }) imaging is also available for our three galaxies. NGC 4244 and IC 2233 have previously been observed using the Wide Field Camera (WFC) of the Advanced Camera for Surveys (ACS) with the $F606W$ filter \citep{seth_study_2005}. These data were accessed from the Multimission Archive at STScI (MAST)\footnote{http://archive.stsci.edu/index.html}. Imaging of UGC 7321 is available from the Wide Field Planetary Camera 2 (WFPC2) and we use a composite $F702W$ and $F814W$ image previously presented in \citet{matthews_modeling_2001}. HST images of the inner regions of each galaxy can be seen in Figure \ref{HST}.

The MIR and FIR observations of our three galaxies were obtained with the {\it Spitzer} Space Telescope's IRAC and MIPS imagers. Data were obtained for this project under {\it Spitzer} PID $20432$ (PI: L. Matthews) along with observations taken from PIDs $3$ (PI: G. Fazio) and $40204$ (PI:R. Kennicutt). These data have recently been used as part of the  {\it Spitzer} Local Volume Legacy (LVL) survey \citep{dale_Spitzer_2009}. This provides imaging and photometry for a large sample of galaxies within $\sim$11Mpc. Data are available for all four IRAC bands ($3.6, 4.5, 5.8 $ and $ 8.0 \mu$m) as well as three MIPS bands ($24, 70$ and $160\mu$m) with photometry provided for each band after foreground stars and background galaxies have been removed and aperture corrections applied. 

Photometry is also available based on near-infrared (NIR) 2MASS\footnote{http://www.ipac.caltech.edu/2mass/} imaging using apertures matched to those used in the extraction of the IRAC and MIPS fluxes. {\it Spitzer} LVL survey imaging can be accessed via an online archive\footnote{http://irsa.ipac.caltech.edu/data/Spitzer/LVL/} and NIR, IRAC and MIPS fluxes are available in Table 2 of \citet{dale_Spitzer_2009}.

 \citet{lee_galex_2011} provide a catalogue of matched far and near-UV observations for galaxies contained in the {\it Spitzer} LVL survey that have been observed with {\it GALEX}. The {\it GALEX} bands are centered at $1528$\AA{} (far-UV) and $2271$\AA{}  (near-UV) \citep{morrissey_-orbit_2005} and fluxes have been extracted using the apertures obtained from the FIR {\it Spitzer} data for consistency. The UV data provide a constraint on the emission from the young stellar population which, due to the higher opacity at UV wavelengths, will play an important role in heating the diffuse dust.  

\begin{table*}
\begin{center}
\caption{Galaxy Properties}
\begin{tabular}{l c c c c c c c }
\hline
Name&Type&$B$ band&Distance&$V_{\rm rot}$&$\mu_{B,i}(0)$\tablenotemark{a}&$M_{HI}$&$M_{H_{2}}$ \\
& &Luminosity& & & & & \\
 & &($10^{9}L_{\odot})$&(Mpc)&(km$s^{-1}$)&(mag$/''^{2}$)&($10^{9}M_{\odot}$)&($10^{7}M_{\odot}$) \\
 \hline
 UGC7321&Sd&$1.1$&$10$&$105$&$23.6$&$1.1$&$3.2$\\
 IC2233&Sd&$1.8$&$10$&$88$&$22.6$&$1.1$&$<0.1$\\
 NGC4244&Sd&$3.6$&$4.4$&$95$&$23.6$&$1.3$&$1.4$\\
\hline 
\label{prop}
\tablenotetext{1}{value deprojected to face-on}
\end{tabular}
\end{center}
\end{table*}

\section{Radiative Transfer Model}\label{code}  
In order to produce synthetic images and SEDs for comparison to the data we adopt a Monte Carlo radiation transfer modelling scheme. 
\subsection{Emissivity and Dust Distributions}
The model allows energy packets emitted within a smooth stellar emissivity distribution to be tracked as they propagate through a dusty medium. For both the emissivity and dust we adopt a smooth ``double exponential'' distribution: 
\begin{equation}\label{double} \rho(\varpi,z) \propto \exp(-|z|/Z)\exp(-{}\varpi/R), \end{equation} 
where $\rho(\varpi,z)$ is the density at a point in the galaxy specified by $\varpi$, the cylindrical radius, and $|z|$ the height above or below the galactic plane. The scale lengths $R$ and $Z$ can be independently varied for both components as necessary. The stellar emissivity and dust density are discretised across a three dimensional spherical grid of cells to allow our Monte Carlo radiation transfer routines to be used. To reduce the number of free parameters in the model no attempt has been made to take account of clumpy substructure within either the emissivity or dust distributions, which may be important in reproducing the optical properties of galactic disks \citep{matthews_modeling_2001,bianchi_effects_2000,pierini_dust_2004}. This also allows the use of a lower resolution grid of cells, reducing the memory requirements of each radiation transfer simulation.

\subsubsection{Emission Sources}\label{emission}
\begin{figure}
\centering
\plotone{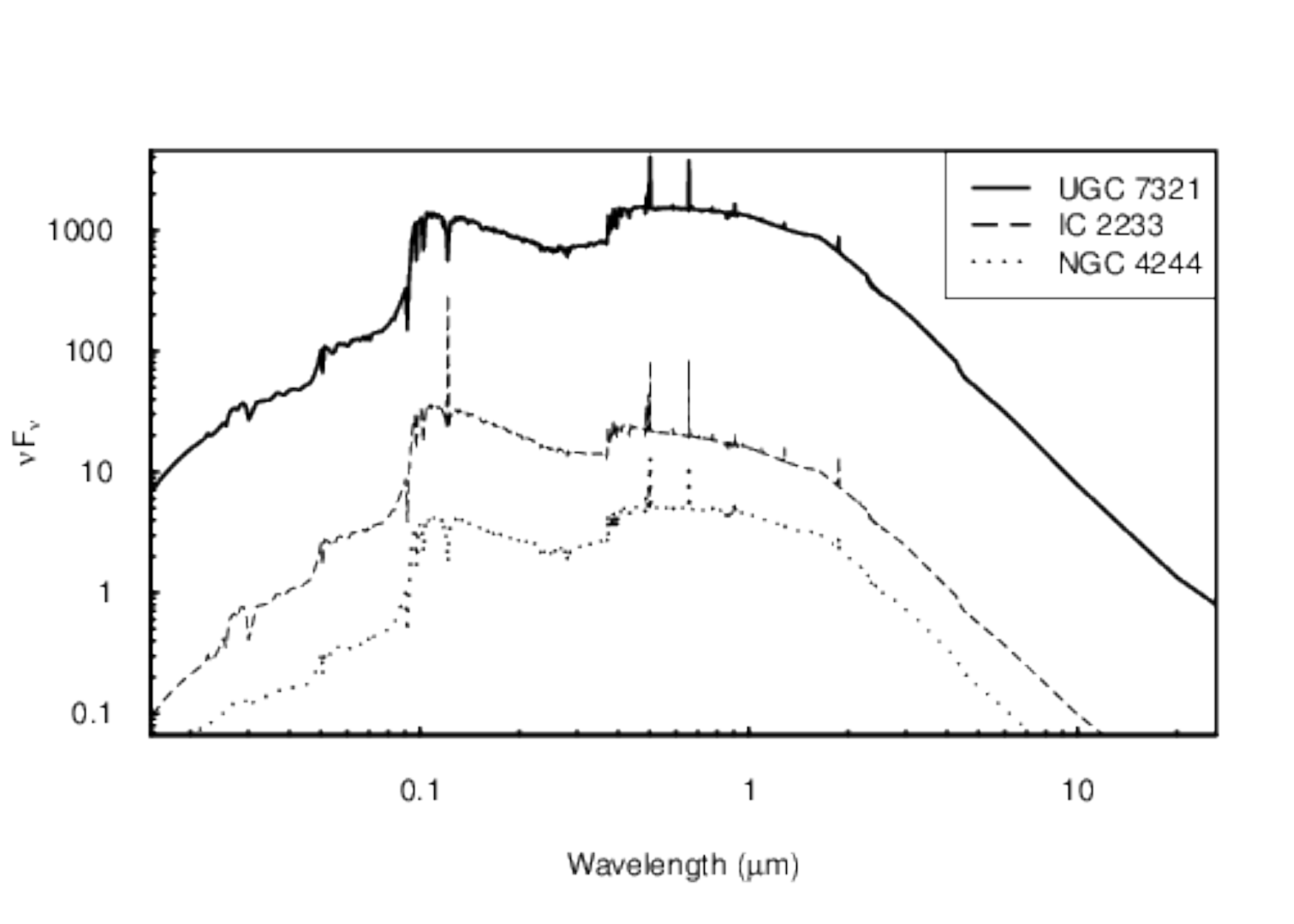}
\caption{The intrinsic stellar emission templates used for our three galaxies. The fluxes have been scaled for comparison. The important property of each template is its shape, as this is used within our models to correctly sample the stellar emission. Each galaxy required a different intrinsic stellar template to match the observation in the UV, optical and NIR. The templates were produced using GALEV \citep{kotulla_GALEV_2009} and model parameters can be found in Table \ref{galev}.  }
\label{stellar}
\end{figure}

To begin the radiative transfer process it is necessary to sample an initial wavelength of emission for each photon packet. Wavelengths are chosen to reproduce the intrinsic unattenuated stellar SEDs produced by the population of stars present in each galaxy. For this purpose a set of synthetic stellar SEDs was constructed using the GALEV\footnote{http://www.galev.org/} codes \citep{kotulla_GALEV_2009}. These are shown in Figure \ref{stellar}. Each model initially has a constant SFR for several Gyrs which is followed by an exponentially decreasing burst of star formation with an $e$-folding time of $1$ Gyr. Table \ref{galev} lists the values used within the model for each galaxy. In all cases a Salpeter initial mass function with slope $\alpha =-2.35$ is used and the metallicity of the gas is allowed to change with time to remain chemically consistent. We also include the effects of continuum and line emission from gas within the model SEDs.

In searching for appropriate stellar populations to model the observed stellar SED we have found that a wide range of ages can provide adequate fits. This has previously been noted by other authors \citep{zackrisson_stellar_2005, vorobyov_age_2009} who have suggested that based on additional information, such as $H\alpha{}$ equivalent widths and oxygen abundance gradients, that old stellar populations are more probable in samples of blue LSB galaxies. The ages adopted here range from $8.6 - 9.4$ Gyrs which are in agreement with \citet{jimenez_galaxy_1998} who suggest ages larger than $7$ Gyrs for LSB galaxies. We note, however, that our conclusions on the dust masses and distributions are most sensitive to the shape of the intrinsic stellar SED and will not be significantly altered by the exact age, metallicity or star formation history of the underlying stellar population which produces it.
\begin{table*}
\begin{center}
\caption{Intrinsic Stellar Template Model Parameters}
\begin{tabular}{l c c c c c }
\hline
Name& Initial SFR&Burst Mass\tablenotemark{a}&Burst Time\tablenotemark{b}&Age\tablenotemark{c}&SFR\tablenotemark{d}\\
 & $M_{\odot}$yr$^{-1}$&Fraction &$10^{9}$ yrs&$10^{9}$ yrs&$M_{\odot}$yr$^{-1}$\\
 \hline
UGC7321&0.02&0.1&7&8.6&0.02\\
IC2233&0.005&0.1&8&9.1&0.04\\
NGC4244&0.03&0.2&7.5&9.4&0.04\\
\hline 
\label{galev}
\tablenotetext{1}{Fraction of remaining gas converted to stars during burst}
\tablenotetext{2}{Age of galaxy at the time of the star formation burst}
\tablenotetext{3}{Age of the galaxy adopted to fit the broadband fluxes}
\tablenotetext{4}{SFR at adopted age}
\end{tabular}
\end{center}
\end{table*}

The SFRs for UGC 7321 and IC 2233 have previously been estimated based on the observed $H\alpha{}$ emission. SFRs of $\sim0.02$ and $0.05$ M$_{\odot}yr^{-1}$ have been found for UGC 7321 \citep{matthews_high-latitude_2003} and IC2233 \citep{matthews_H_2008} respectively, which are in reasonable agreement with the values in Table \ref{galev}. We have estimated the SFR of NGC 4244 based on the $H\alpha$ luminosity, $L_{H\alpha} = 10^{40}$erg s$^{-1}$ \citep{kennicutt_h_2008}, and the SFR - $H\alpha$ relation taken from \citet{kennicutt_star_1998}. Together these yield a SFR $\sim 0.079 \pm 0.023$ M$_{\odot}yr^{-1}$ which is roughly consistent with the value adopted for the template stellar SED.

A single stellar SED is assumed for each galaxy modeled. No account of stellar gradients, which may be important in LSB galaxies \citep{matthews_modeling_2001} is made in order to reduce the complexity of the model. We do not believe that the stellar gradients will have a significant impact on the SED and derived dust disk properties.

\begin{figure}
\centering
\plotone{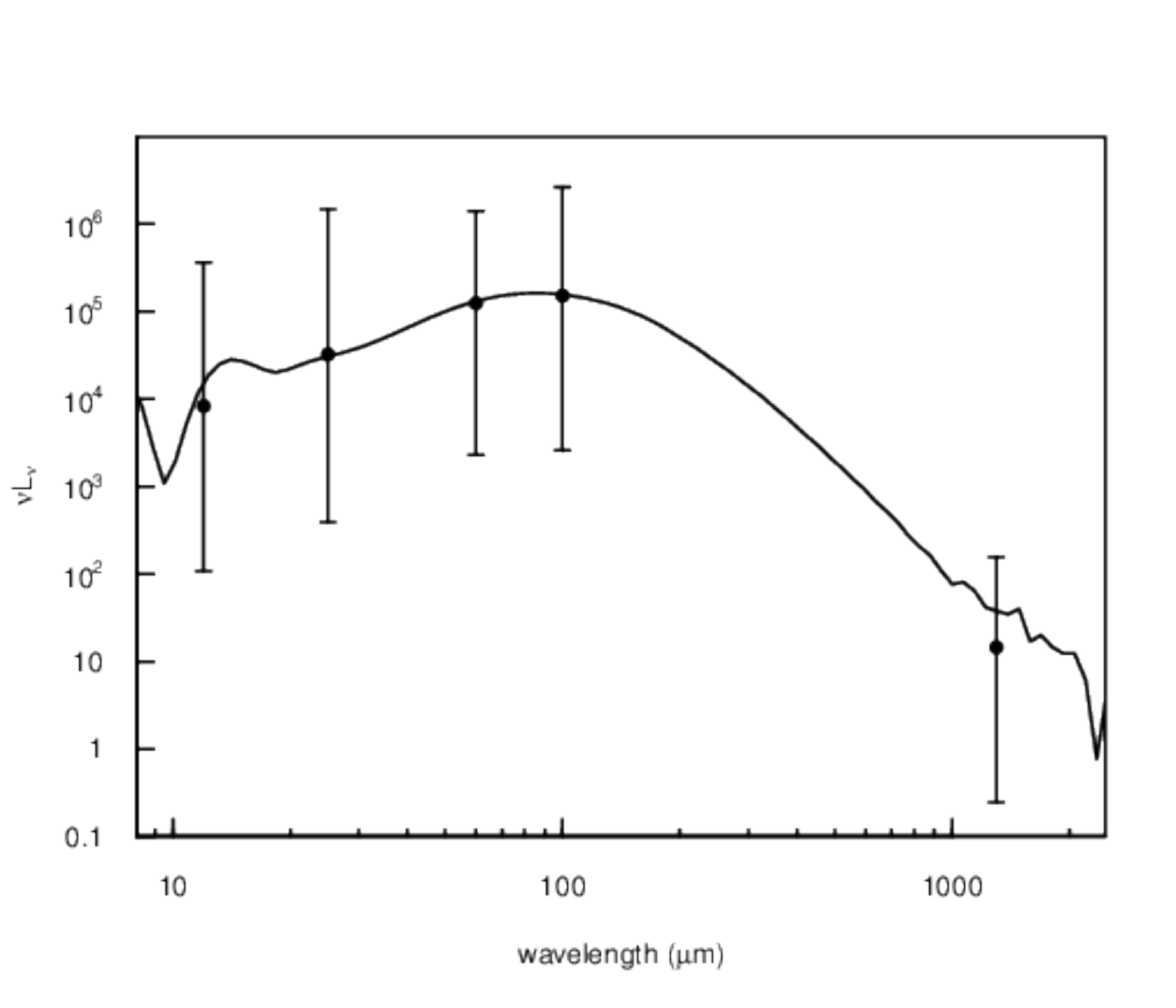}
\caption{The template for the emission from compact sources between $10$ and $2000 \mu$m. The solid line shows the model output SED as it varies with wavelength. The model comprises of a central star (T=36,000K) heating a constant density shell of gas and dust surrounding it. The shell extends from $10^{-4} - 2.3$pc and has a V band optical depth of $\tau_{V} \sim 88$. The total luminosity included in each cloud model is $2.5 \times10^{5} L_{\odot}$ and each has a dust mass of $5 \times10^{4} M_{\odot}$. Data points represent the median observed properties of 56 compact \HII{} regions found by \citet{chini_1.3_1986} at $12,25,60,100$ and $1300 \mu$m. Error bars show the range of values observed at each wavelength.  }
\label{compact}
\end{figure}

Also included is a secondary emission component from obscured star formation regions that may be necessary to account for the total FIR emission. As such emission takes place from compact regions that are below the resolution of the spherical grid currently employed in the radiation transfer, a separate model has been created. The regions are approximated by a single central star surrounded by a shell of dust and gas. The shell has a $V$ band optical depth of $\tau_{V} \sim 88$ and extends from a distance $d=10^{-4} $pc out to $2.3$pc. The size of the shell was chosen to best reproduce the shape of the observed FIR emission of star forming regions observed by \citet{chini_1.3_1986}. The shell is exposed to internal heating from a stellar source, with emissivity sampled from a model atmosphere \citep{kurucz_kurucz_1993} of temperature T $=36,000$K. The shell is also illuminated from the exterior by a radiation field approximating the mean intensity in the plane of the galaxy model. Due to the size of the clouds, however, the luminosity incident on the exterior is typically much smaller than the luminosity of the central source and hence diffuse illumination has little effect on the overall SED.  The output SED can be seen in Figure \ref{compact} along with observed data points of 56 Galactic compact \HII{} regions taken from \citet{chini_1.3_1986}. A single SED is used to approximate all compact sources in the galaxy. The spatial distribution of compact sources is an exponential disk (Eq. \ref{double}) with $Z=0.1$ kpc and $R=1.8$ kpc in all models. This secondary emission component, from obscured star forming regions, has previously been found necessary to reproduce the FIR emission in HSB galaxies \citep{popescu_modelling_2000,misiriotis_modeling_2001,popescu_modelling_2011}. Other methods to account for such emission include utilizing sub grid resolution models for individual star forming regions \citep{jonsson_high-resolution_2010} or the inclusion of adaptive grids to allow radiation transfer on the small size scales needed. Current adaptive grids models, however, lack the resolution to resolve individual star forming regions \citep{bianchi_dust_2008}.

\subsubsection{Dust Model}\label{dustmodel}
The dust content of the model galaxy is treated in a two component manner as described by \citet{wood_emission_2008} in order to take account of both large grains and small transiently heated grains. Large grains with sizes $>200\textrm{\AA }$ are assumed to have attained radiative equilibrium temperatures. In the case of these grains the \citet{bjorkman_radiative_2001} method is used to update the temperature of the cell and correctly sample the emissivity for the reemitted photon packet when an interaction takes place. The advantage of this method lies in the way the frequency of the reemitted photon is sampled to correct the frequency distribution of the photons previously emitted from the cell to the new cell temperature. The SED of the cell then relaxes to the equilibrium value, as the simulation progresses, without the need for iteration. The composition and size distribution of the large grains is taken from the work of \citet{kim_size_1994} who used a mixture of silicate and graphite grains to fit the observed extinction in the Milky Way. This treatment of large dust grains, without the extensions added to treat transient heating of very small grains (VSG) and polycyclic aromatic hydrocarbon (PAH) molecules, has previously been used by \citet{savoy_scuba_2009} to investigate the dust content of early-type galaxies.

The \citet{bjorkman_radiative_2001} method cannot, however, be applied to transiently heated VSGs and PAH molecules that are not at equilibrium temperatures. Emission from these grains is important in reproducing the MIR emission. For photon packet interactions with grains that are smaller than $200\textrm{\AA }$ the methods introduced in \citet{wood_emission_2008} are used. The emission from PAH molecules and VSGs is calculated from precomputed emissivity files from \citet{draine_infrared_2007}. The emissivity is assumed to vary only with the mean intensity of the radiation field and not with the spectral shape. The cutoff in the grain size at  $200\textrm{\AA }$ is somewhat arbitrary but it is thought that grains above this will normally reach equilibrium temperatures \citep{draine_infrared_2007}. In order to correctly sample the emissivity of the PAH molecules and VSGs a value for the mean intensity in the cell where the interaction took place is needed. This requires an iterative process where initially a value of $J=1$ is used, $J$ being the value of the mean intensity relative to the \citet{mathis_interstellar_1983} value of the local interstellar radiation field (ISRF) of $J_{ISRF}=2.17\times 10^{-2}$ erg cm$^{-2}$ s$^{-1}$. Consecutive iterations then use the mean intensity computed for each cell on the previous iteration to calculate the emissivity of the PAH molecules and VSGs. Due to the small contribution of photon packets reprocessed by PAH molecules and VSGs to the total mean intensity it is found that typically three iterations are required \citep{wood_emission_2008}.

As the ISM conditions of LSB galaxies are likely different from those found in the Milky Way, it is possible that our treatment of the PAH/VSG dust components may be inappropriate. It is known that the strength of the PAH emission relative to that from VSGs, estimated using the $f_{\nu}$($8\mu$m)/$f_{\nu}$($24\mu$m) colour, decreases with decreasing metallicity \citep{engelbracht_metallicity_2005,madden_ism_2006}. This is likely due to either the delayed formation of PAH molecules in low metallicity environments \citep{galliano_stellar_2008} or destruction/processing caused by the harder radiation field \citep{gordon_behavior_2008,madden_ism_2006}. It is therefore possible that our treatment, which assumes Milky Way abundances, may overestimate PAH emission relative to that of the VSGs. The fraction of the total dust mass contributed by PAH/VSGs is also fixed to a value calculated for the Milky Way. This could affect the relative strength of the MIR and FIR emission.  

To model the appearance and SED each galaxy we can alter the following free parameters of the model: stellar scale length, stellar scale height, dust scale length, dust scale height, stellar luminosity, total dust mass, inclination and the luminosity of the obscured star formation component.

The overall fitting process was done manually to find the set of parameters that could provide adequate fits to both the optical and the UV-FIR SED. 

\section{Results}\label{results}
Initial values for the fitting process were taken from sources in the literature. \citet{matthews_modeling_2001} have previously investigated UGC 7321 using a three dimensional Monte Carlo scattered light code and their parameters for a smooth stellar and dust distribution were adopted as initial values in this case. \citet{matthews_H_2008} and \citet{fry_deep_1999} have previously estimated the scale lengths of IC2233 and NGC 4244 respectively, by fitting an exponential function to the observed surface brightness profiles. While the functions used take no account of the effects of dust on the galaxy profiles they should provide reasonable initial estimates for the stellar distributions assuming that dust effects are not severe. From the work of \citet{xilouris_are_1999} on high surface brightness, edge-on, galaxies we initially adopted the relation that the dust scale heights would be approximately half the stellar scale heights and the dust scale lengths around $1.4$ times larger than the stellar values. 

One of the main difficulties in modeling the dust distributions of LSB galaxies is that their disks appear to be optically thin, even when viewed edge on. When modeling the appearance of HSB galaxies, the effects of the dust distribution can be quantified based on the presence of a dust lane along the galactic mid-plane. As none of our galaxy sample shows any sign of a dust lane the problem becomes more degenerate. Initially we increased the dust mass in the adopted geometry until the effects of the dust became obvious and the optical profiles of the model galaxy no longer matched the shape of the observed data. It was found that the dust mass required to reprocess the stellar light and reproduce the observed FIR emission could not be located in an average HSB type dust disk without revealing its presence by flattening the optical profile in the central region of the galaxy.
\begin{figure*}
\centering
\plotone{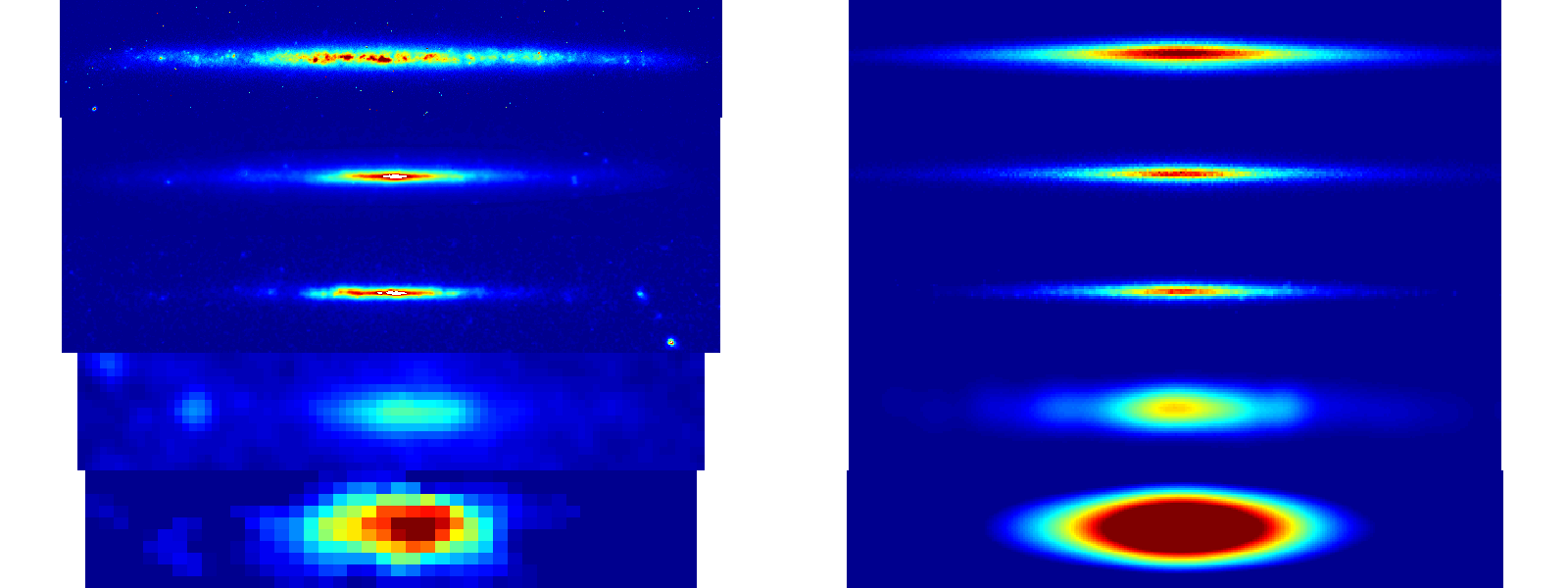}
\caption{Observed (left) and synthetic (right) images of UGC 7321 at, from top to bottom: $B$ band, $3.6\mu$m, $8\mu$m, $70\mu$m and $160\mu$m. }
\label{ugc_7321_im}
\end{figure*}
\begin{figure*}
\centering
\plotone{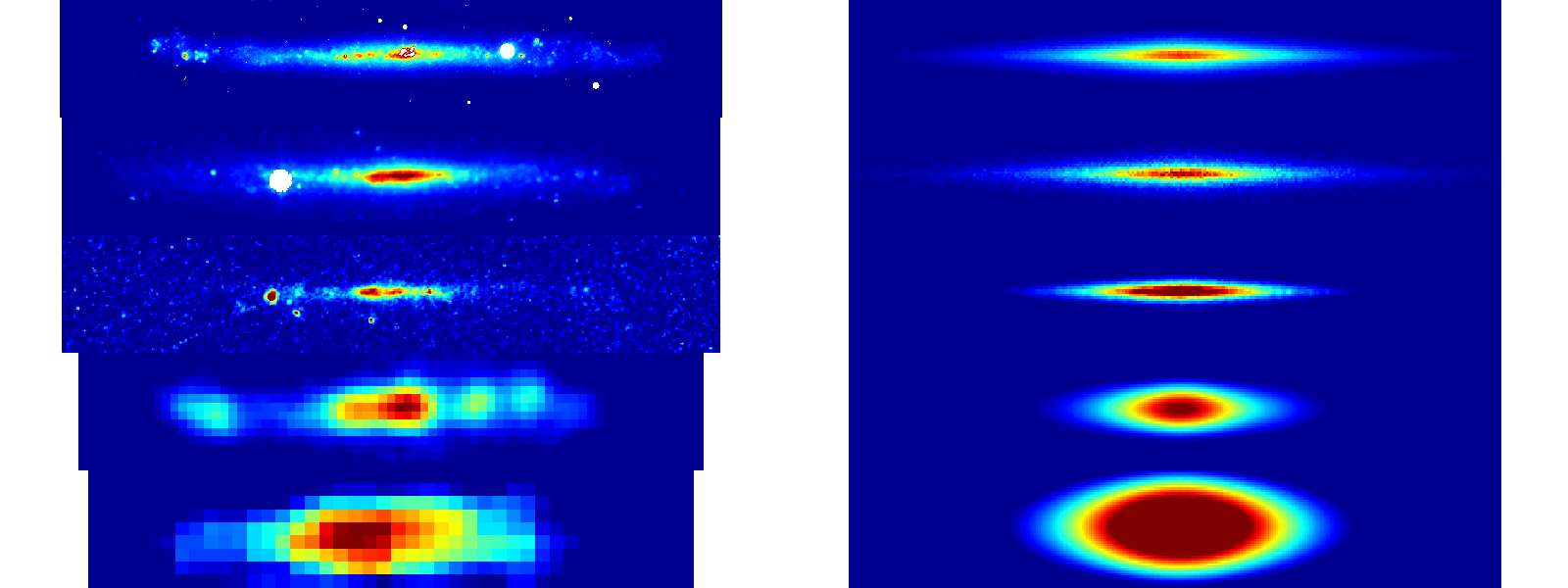}
\caption{Observed (left) and synthetic (right) images of IC 2233 at, from top to bottom: $B$ band, $3.6\mu$m, $8\mu$m, $70\mu$m and $160\mu$m. }
\label{ic2233_im}
\end{figure*}
\begin{figure*}
\centering
\plotone{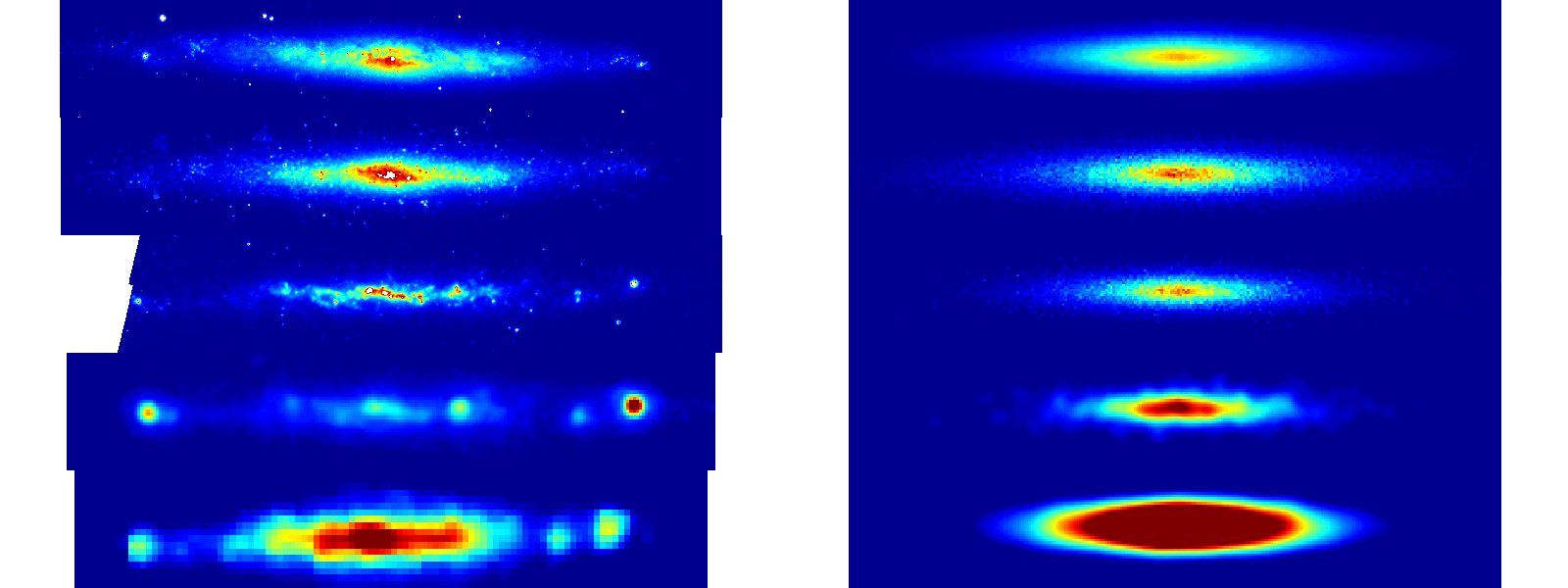}
\caption{Observed (left) and synthetic (right) images of NGC 4244 at, from top to bottom: $r$ band, $3.6\mu$m band, $8\mu$m, $70\mu$m and $160\mu$m.}
\label{ngc4244_im}
\end{figure*}

We find that the optical appearance and total FIR emission of the LSB galaxies can be reproduced by removing the constraint that the dust be more concentrated in the galactic midplane than the stars. If the dust scale height is taken to be equal to the stellar scale height then it is also possible to fit the optical appearance of the galaxy adequately. In this case, however, a significantly larger dust mass can be accommodated within the dust disk without the appearance of a dust lane. In order to reproduce the observed FIR flux distribution it was also necessary to increase the dust scale length. When the dust is vertically extended into a disk with a larger scale height that is well mixed with the stellar population, the exact radial distribution of the dust becomes poorly constrained from optical data alone. We have found the dust scale length is between $1.8-2.6$ times the stellar scale length for our LSB disk galaxies.

The best fit parameters for our models of each galaxy can be found in Table \ref{fit}. $R_{s}$ and $R_{d}$ are the radial scale lengths of the stars and dust while $Z_{s}$ and $Z_{d}$ are the vertical scale heights. $L_{s}$ is the intrinsic bolometric luminosity of the stellar population. $M_{d}$ represents the total dust mass and $i$ the inclination angle. $L_{cloud}$ gives the luminosity emitted from the template of compact dust emission (see section \ref{emission}). $\tau_{face}$ and $\tau_{eq}$ give the $V$ band face-on and equatorial optical depths respectively\footnote{$\tau_{face}$ is the optical depth from the disk centre to the edge of the model grid perpendicular to the plane of the disk. $\tau_{eq}$ is measured from the disk center to the edge of the model grid parallel to the disk plane.}.

\begin{figure}
\centering
\includegraphics[width=240pt]{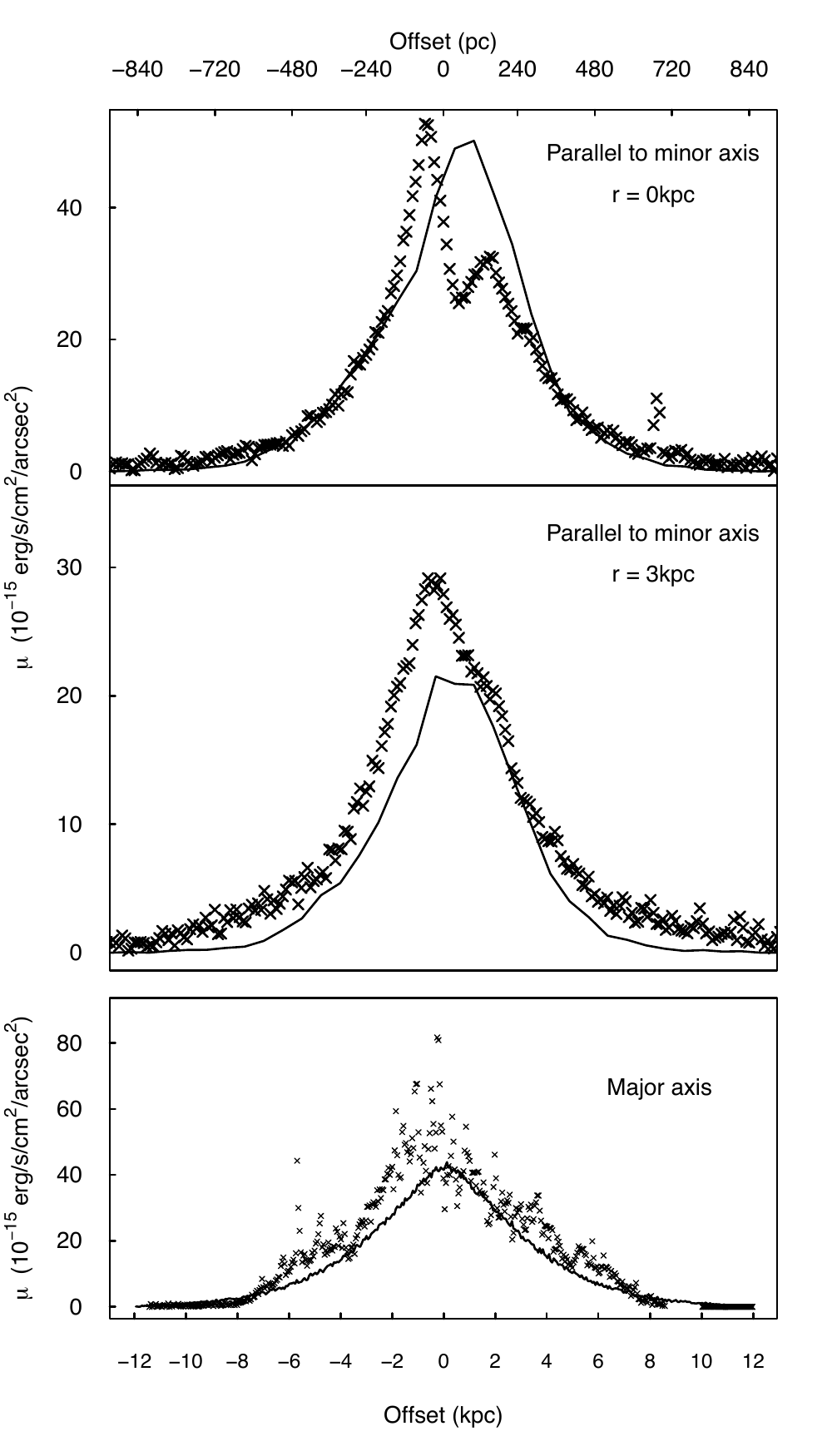}
\caption{The top two plots show UGC 7321 $B$ band surface brightness slices parallel to the minor axis at the galaxy centre (top), and 3 kpc radial distance along the major axis (middle). The lower panel shows the surface brightness variation along the major axis. All slices are average values taken across a $\sim 60$pc region. Solid lines show the predicted intensity based on our best fitting Monte Carlo radiation transfer models and crosses represent the data taken from the $B$ band image.}
\label{7321_slice}
\end{figure}

\begin{figure}
\centering
\includegraphics[width=240pt]{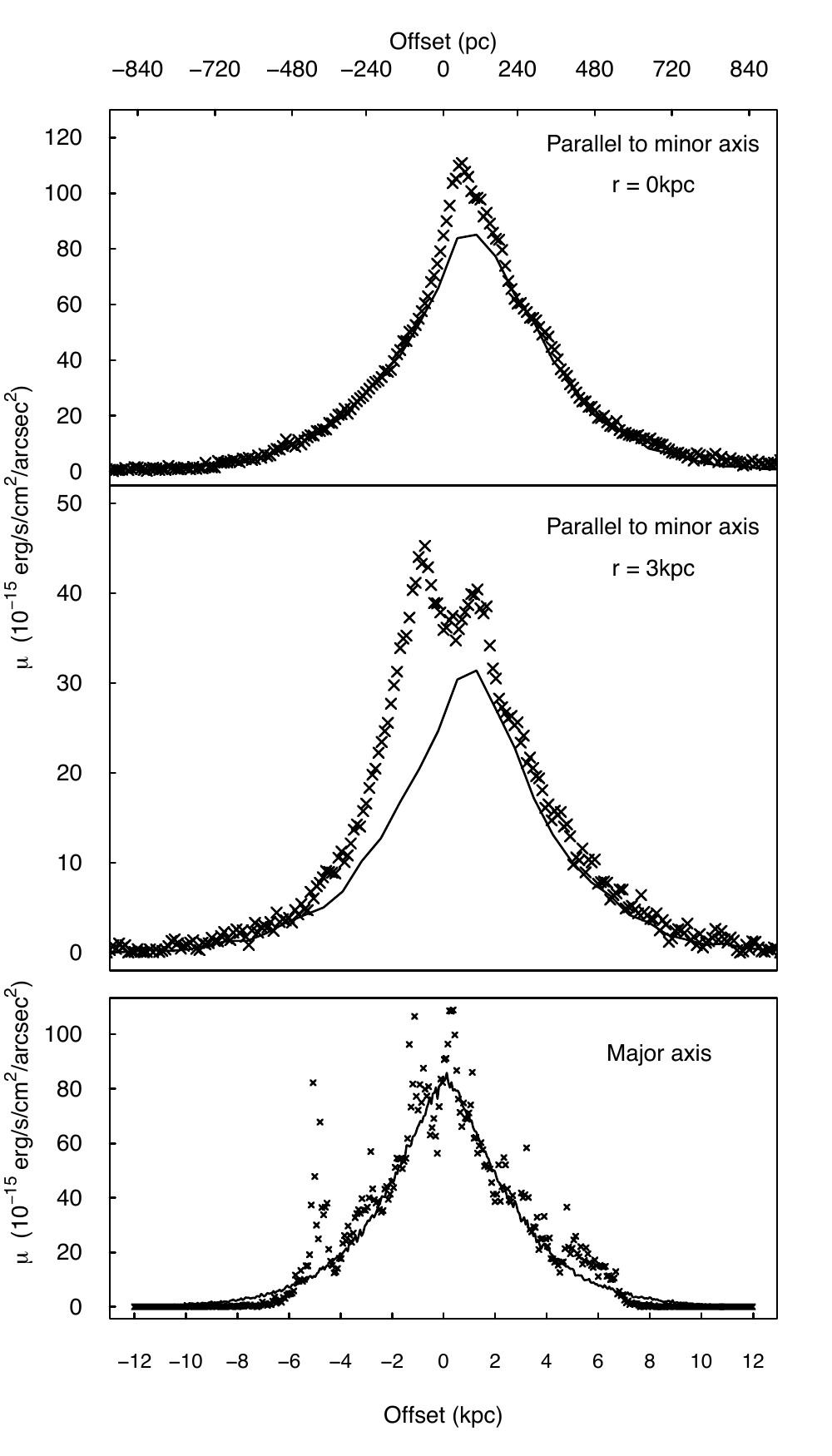}
\caption{As in Fig. \ref{7321_slice}, but for IC 2233.}
\label{ic2233_slice}
\end{figure}

\begin{figure}
\centering
\includegraphics[width=240pt]{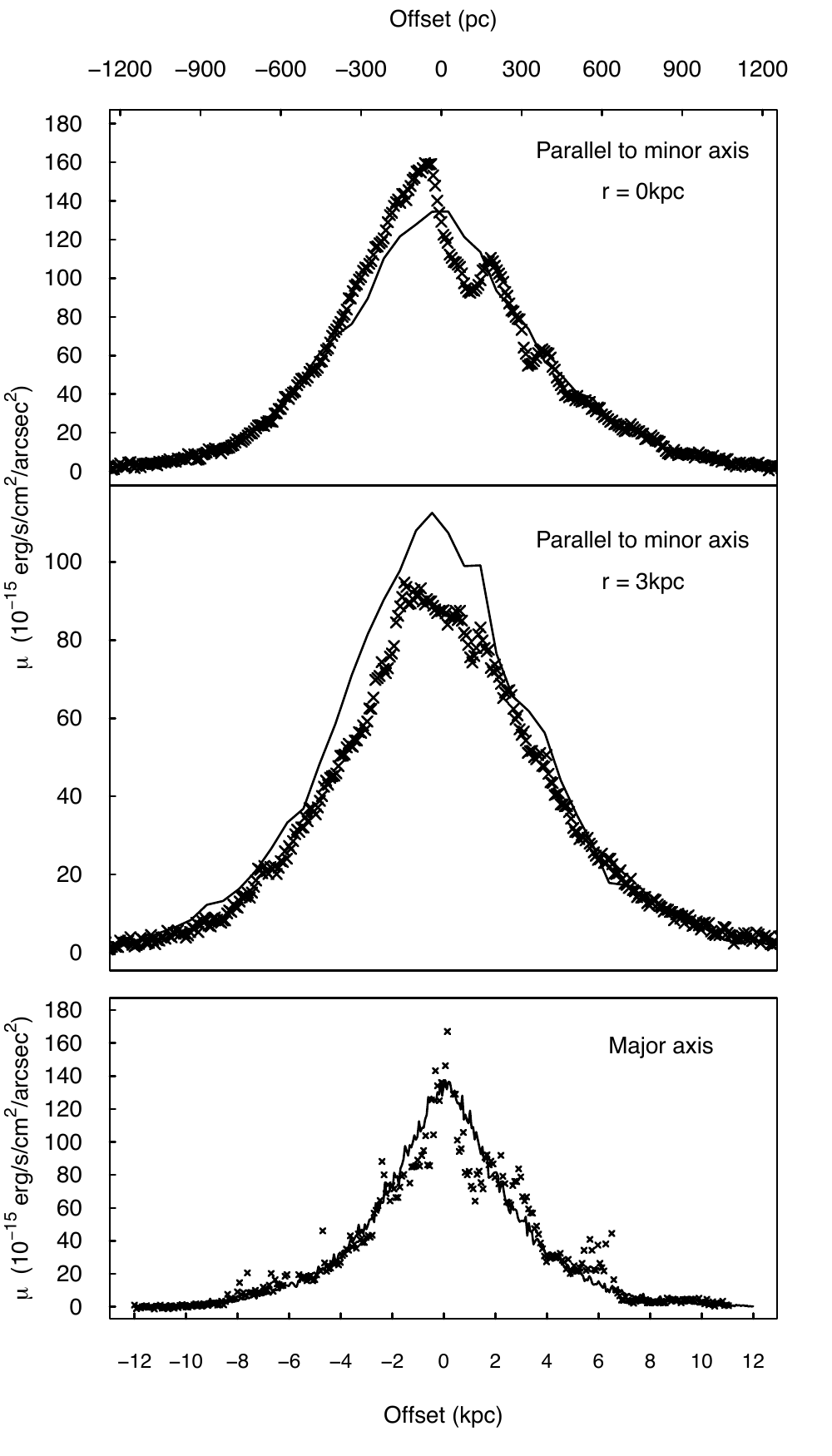}
\caption{As in Fig. \ref{7321_slice}, but for NGC 4244 and utilizing SDSS $r$ band rather than $B$ band data.}
\label{ngc4244_slice}
\end{figure}

\begin{figure*}
\centering
\leavevmode
\includegraphics*[width=450pt]{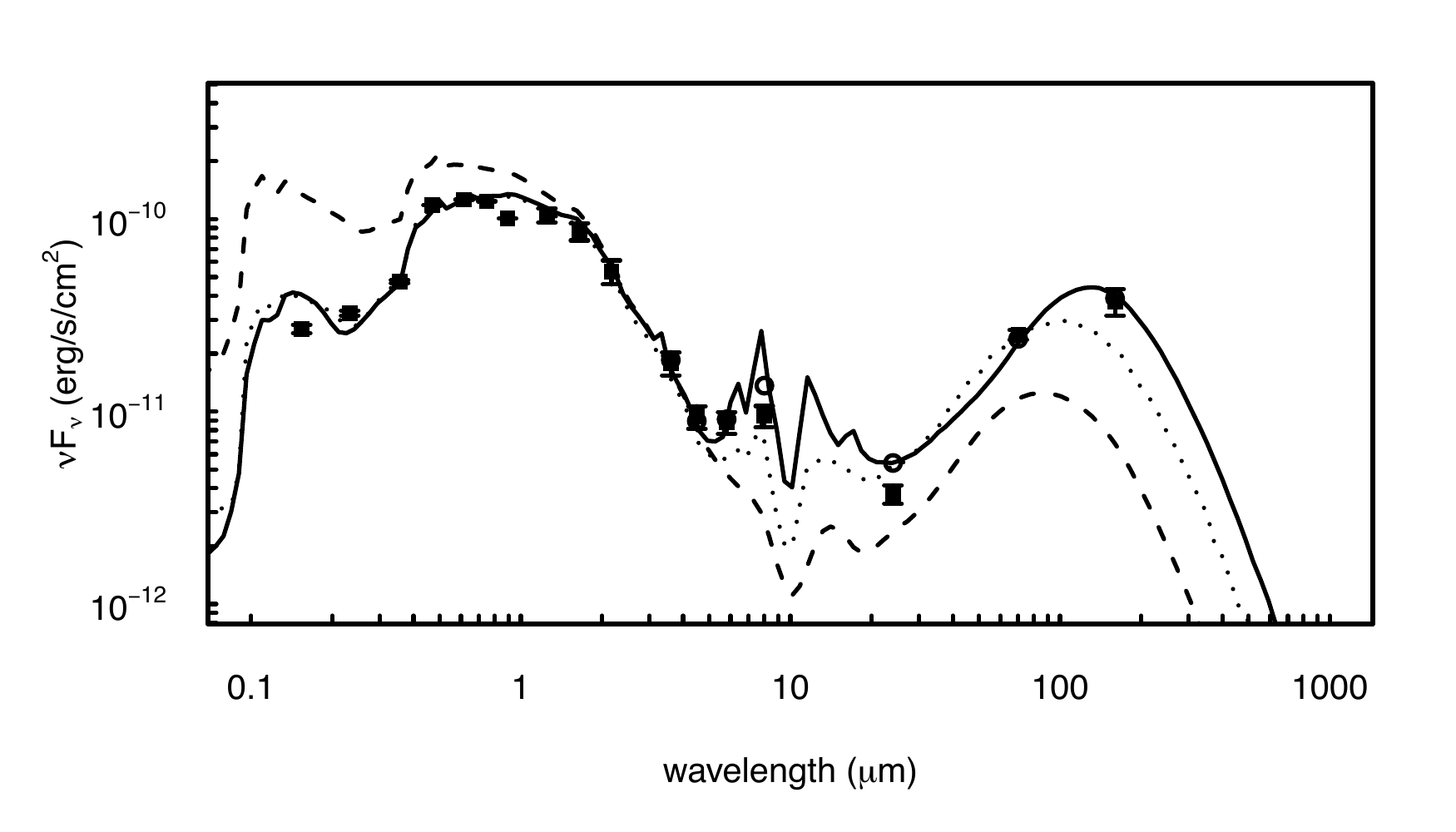}
\caption{The observed and modeled SED of UGC 7321. Square points indicate an observed flux from one of our data sources either {\it GALEX}, SDSS, 2MASS or {\it Spitzer} IRAC/MIPS along with their associated errors. Error bars associated with SDSS imaging do not include the contribution of the sky subtraction errors. The solid line shows the model output SED as calculated from the best-fit to the WIYN B band image. Open circles show the predicted fluxes for {\it Spitzer} IRAC/MIPS bands. The long dashed line indicates the input, unattenuated, stellar template plus compact dust emission template. The dotted line shows the SED from the best-fit model when an average stellar-dust distribution for HSB galaxies \citep{xilouris_are_1999} is used to fit the optical imaging.}
\label{7321_sed}
\end{figure*}

\begin{figure*}
\centering
\leavevmode
\includegraphics*[width=450pt]{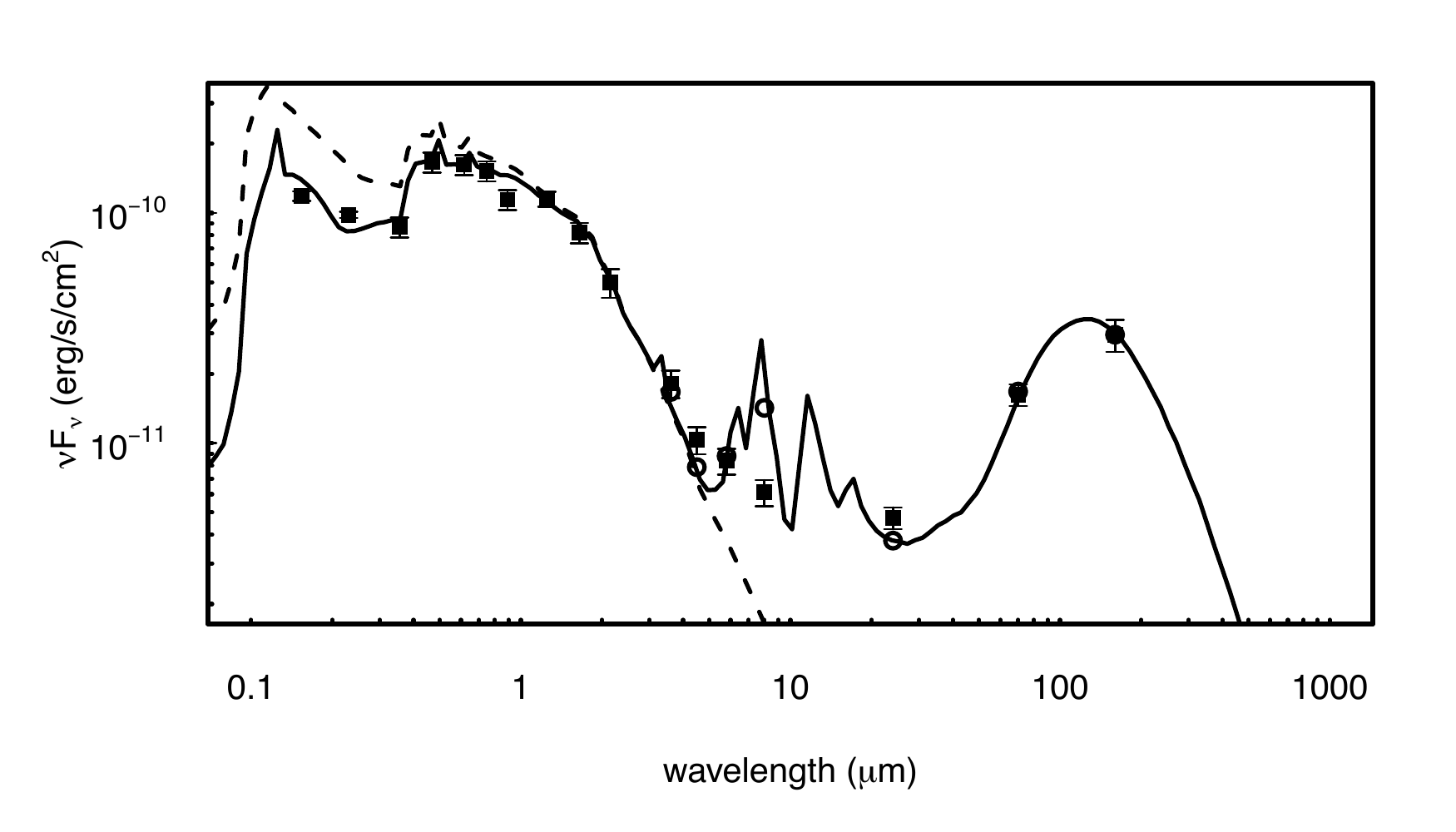}
\caption{The above plot shows the model SED of IC 2233 (solid line) along with the available data (squares). Open circles indicate the predicted model fluxes for the filters used in the observations. The dashed line shows the intrinsic SED of the emissivity sources (both stellar and dust emission from unresolved sources).  }
\label{ic2233_sed}
\end{figure*}

\begin{figure*}
\centering
\includegraphics[width=450pt]{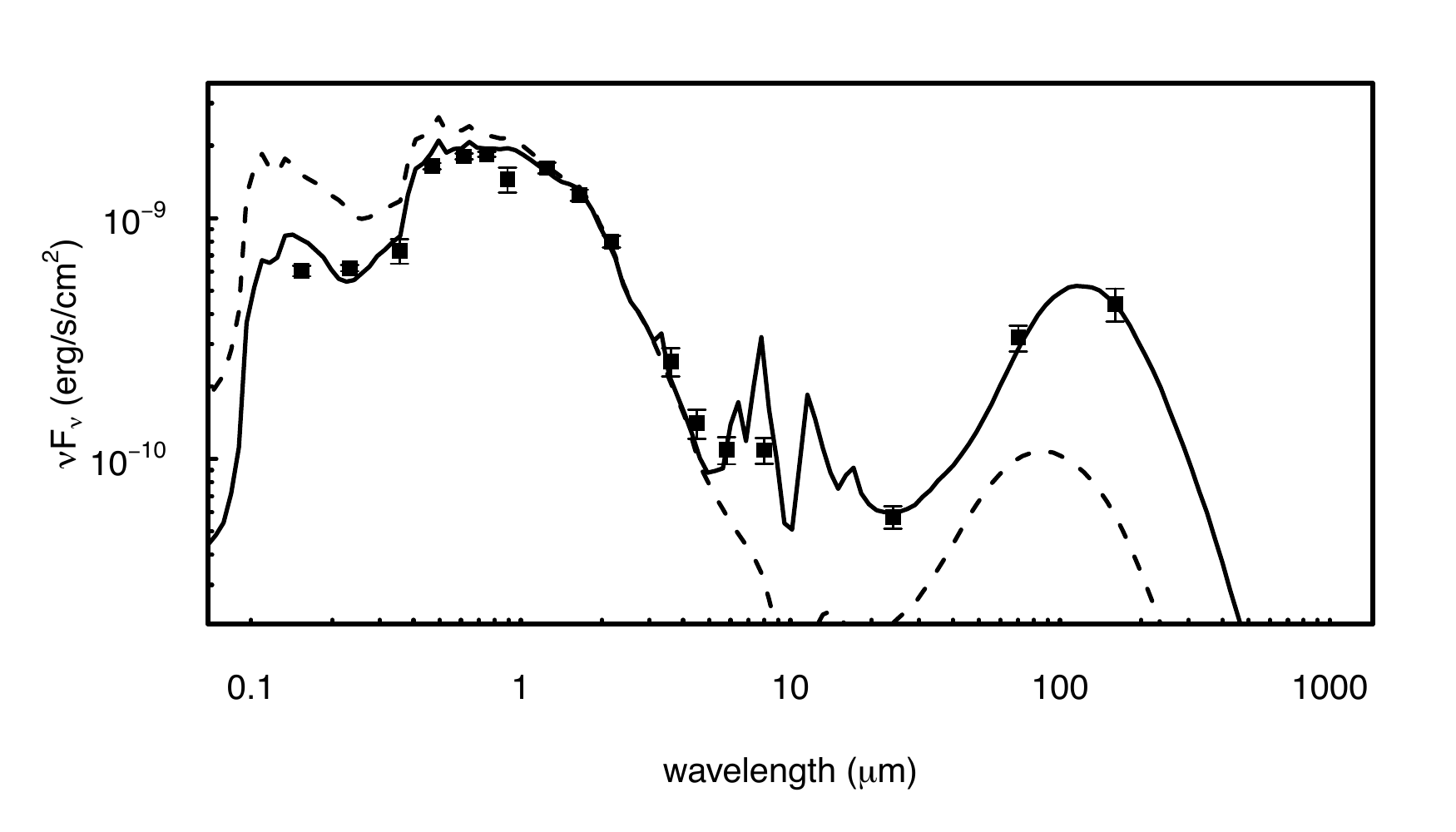}
\caption{Same as Figure \ref{ic2233_sed} but for NGC 4244.}
\label{ngc4244_sed}
\end{figure*}

Figures \ref{ugc_7321_im}, \ref{ic2233_im} and \ref{ngc4244_im} show a comparison of the $B$ (or $r$), $3.6\mu$m , $8\mu$m, $70\mu$m and $160\mu$m data (left panel) and synthetic images (right panel). In general the  $B$ (or $r$) and $3.6\mu$m model images match the large scale morphology of the data reasonably well. Differences are, however, apparent in the longer wavelength emission at $8$, $70$ and $160\mu$m that is dominated by emission from PAHs and dust grains. IC 2233 and NGC 4244 show significant structure in their $8$ and $70\mu$m images, suggesting that perhaps a more complex non-axisymmetric dust distribution is needed. In both galaxies several small point like sources in the outer disk are prominent sources at $70\mu$m. These are likely star forming regions as they appear brighter at shorter wavelengths and show associated emission in {\it GALEX} far-UV and H$\alpha$ imaging. The $160\mu$m emission from all three of our model galaxies shows a significant deviation from the data. The models predict a centrally concentrated $160\mu$m image while the data suggest a more diffuse distribution. Figures \ref{7321_slice}, \ref{ic2233_slice} and \ref{ngc4244_slice} show $B$ (or $r$) band intensity profiles parallel to the minor axes of the galaxies at various points while figures \ref{7321_sed}, \ref{ic2233_sed} and \ref{ngc4244_sed} show the output model SEDs for the three LSB galaxies.

\begin{table*}
\begin{center}
\caption{Fitted Galaxy Parameters}
\begin{tabular}{l c c c c c c c c c c}
\hline
Name&$R_{s}$&$Z_{s}$&$R_{d}$&$Z_{d}$&$L_{s}$&$M_{d}$&$i$&$L_{cloud}$&$\tau_{face}$&$\tau_{eq}$\\
 &(kpc)&(kpc)&(kpc)&(kpc)&$(10^{8}L_{\odot})$&$(10^{6}M_{\odot})$&$(^{\circ})$&$(10^{8}L_{\odot})$ && \\
\hline
UGC7321&1.8&0.12&4.30&0.12&14.5&2.25&88.5&0.67&0.072&1.99\\
IC2233&1.7&0.18&4.50&0.18&19.3&1.16&88.5&0.01&0.034&0.69\\
NGC4244&1.9&0.2&3.4&0.2&33.5&2.38&84.5&1.0&0.106&1.57\\
\hline
\label{fit}
\end{tabular}
\end{center}
\end{table*}

Overall the models are able to reproduce the global properties of the data. The surface brightness slices parallel to the minor axis and along the major axis show a similarity to the data and deviations are likely caused by the comparison of smooth axisymmetric models to observations of a galaxy that shows a clear clumpy structure (see Figure \ref{HST}). The main discrepancies found between the surface brightness of the model and data generally occur in the central regions of the galaxies. As can be seen in Figure \ref{HST} these are also the regions that show the greatest number of dark clouds and bright star clusters. Our smooth axisymmetric model is not able to reproduce such structures and so only a relatively poor fit can be achieved in the central regions. 
The short dashed line in Figure \ref{7321_sed} shows the result of fitting the optical imaging of UGC 7321 using the average HSB galaxy stellar-to-dust scaling relation of \citet{xilouris_are_1999}. In order to produce the required $70\mu$m emission a large luminosity is assigned to the obscured star formation template (see section \ref{emission}). However, this scenario underestimates the $160\mu$m emission by a factor $\sim 2$.

In the SEDs the shorter wavelength data from the UV through to NIR originate in the underlying stellar populations and can be highly attenuated by the diffuse dust in the ISM. Most of the photons absorbed by the dust, in our models, originated as short wavelength UV emission. This can be seen in Figure \ref{7321_sed} as the difference between the intrinsic stellar emission (long dashed line) and the observed emission (solid line) once the photons have propagated through the dusty ISM. The amount of absorption tends to decrease with increasing wavelength as the dust opacity decreases and the photons are able to travel through the ISM with a lower probability of interaction. Our models can reproduce the observed optical and NIR fluxes well but show slight discrepancies in the FUV emission. The models are also unable to reproduce the $z$ band flux ($0.893 \mu$m). We believe this may be due to the lower S/N in the $z$ band combined with the intrinsic low surface brightness of the galaxies, leading to galaxy flux being lost during removal of the sky background. 

The $5.8$ and $8.0\mu$m IRAC bands trace emission from PAH molecules while the MIPS $24 \mu$m emission is produced primarily by the warm VSGs. In all three cases the $8.0\mu$m emission is over-predicted by our models. The $24\mu$m flux is over-predicted by the model of UGC 7321, under-predicted for IC 2233. This behavior in the MIR is likely a result of our treatment of the PAH/VSG emission which is based on Milky Way abundances and illuminating radiation field shape (see \ref{dustmodel}). Both the relative abundances of PAH molecules and VSGs in LSBs and the shape of the ISRF in LSBs are likely different causing the discrepancies observed in the MIR. 

The larger, cooler dust grains are responsible for the peak of the FIR emission bracketed by the MIPS observations at $70$ and $160\mu$m. In all three cases we have been able to reproduce the FIR $70$ and $160\mu$m emission, within the quoted photometry errors, using our models.

\section{Discussion}\label{discuss}
The model outputs suggest that it is possible to provide a good global fit to the optical appearance of LSB galaxies, while at the same time reproducing the observed FIR emission. At some locations, however, such as the central region of UGC 7321 (Figure \ref{7321_slice}, top) there is a clear discrepancy between the data and model. In most cases this occurs due to structure in the galaxy that is not included in our models, such as
star clusters, \HII{} regions, and optically thick dust clumps.

\subsection{Implications of the Models}
In our analysis we have found that in order to accommodate the mass of dust required to reprocess the FIR emission seen in our sample of three LSB galaxies it is necessary for the dust to be located in a diffuse disk with a scale height at least equal to that of the stellar population. This is in contrast to the dust distributions obtained by previous studies of HSB galaxies (see Section \ref{results}), which find the dust to be located in a diffuse disk with approximately half the scale height of the stellar disk. The main observational signature of the smaller vertical extent of the dust in HSB galaxies is the central dust lane observed when the inclination approaches edge-on. \citet{dalcanton_formation_2004} have previously noted that in a sample of edge on galaxies that spans a range in $V_{\rm rot}$ there appears to be a remarkably sharp divide between galaxies that  contain dust lanes and those that do not at around $V_{\rm rot} =120$kms$^{-1}$, with no abrupt change in the stellar or \HI{} surface densities. They attribute this sharp divide to the onset of radial instabilities in galactic disks causing the vertical collapse of the ISM when perturbed by spiral structure. The collapse leads to a higher gas density in the central galaxy having the effect of increasing the star formation rate and surface brightness. The increased gas density also causes a corresponding increase in the dust density in the galactic mid-plane which, when viewed edge-on, will produce the pronounced dust lanes observed. From Table \ref{prop} it can be seen that all three of our LSB galaxies have $V_{\rm rot} < 120$kms$^{-1}$ and so from the results of \citet{dalcanton_formation_2004} would be expected to be stable across most of their disks. This result suggests that as the disks are stable the gas and dust disk should have a scale height comparable to or larger than the stellar value, which is what we have found from our modeling of the optical, NIR and FIR data.

In the case of the LSB galaxies studied here an alternative explanation for the comparable scale heights of the dust and stellar disks is that the scale height of the stellar disk is comparatively smaller than that found in HSB galaxies, contributing to the thin appearance of the galaxies. In this case the dust and gas disk scale height may be comparable to more massive galaxies but a drop in stellar scale height has concentrated the stars within the dusty ISM. \citet{bizyaev_stellar_2004} have observed that stellar disks appear to become thinner as the $R$ band central surface brightness dims, which provides evidence that we may indeed be seeing a reduced stellar scale height in the LSB galaxies studied here. This effect may be due to the inefficiency of processes that contribute to the vertical heating of the stellar component such as scattering by giant molecular clouds, the effects of bar formation and a lack of gravitational interactions with neighboring galaxies. 

Alternatively we may be observing a combination of both the processes mentioned above. An increase in the dust disk scale height caused by the stability of the ISM, coupled with a relative flattening of the stellar disk. Both effects combined would then result in comparable scale heights of the dust and stellar distributions.

Previous works have suggested that the dust scale length in galaxies may range from being only slightly larger than the stellar scale length \citep{munoz-mateos_radial_2009} to being moderately extended \citep{xilouris_are_1999,tempel_dust-corrected_2010} or even up to an order of magnitude larger \citep{holwerda_opacity_2005}. The average value of dust scale length we find is R$_{d} = 2.3 \pm 0.24~$R$_{s}$ which is slightly larger than previous values based on FIR and optical extinction modeling but significantly smaller than the value estimated by \citet{holwerda_opacity_2005} from background galaxy counts.

The low star formation rates observed in LSB galaxies are likely caused by the vertically and radially extended gas distributions leading to low surface and volume densities that inhibit star formation via the formation of cold, dense clouds. \citet{van_der_hulst_star_1993} find a mean peak \HI{} surface density, in a sample of LSB disk galaxies, of $4.6\pm1.4$ M$_{\odot}$pc$^{-2}$. This value is around a factor $\sim2$ lower than that found by \citet{cayatte_very_1994} for HSB Sd galaxies. Although the low density of LSB galaxy disks leads to gravitational stability and prevents global star formation driven by spiral density waves, star formation still occurs in localized high density regions. H$\alpha$ imaging suggests that this does not occur in the same pattern in all LSB disk galaxies. UGC 7321 shows H$\alpha$ emission concentrated in a thin layer \citep{matthews_extraordinary_1999} while IC 2233 and NGC 4244 show more vertically extended emission and large star forming complexes in the outer disk \citep{matthews_H_2008,hoopes_diffuse_1999}. These differences in the morphology of the current star formation suggest that there may be differences in the clumpy ISM structure. This may indicate a more gradual transition in the ISM structure across galaxies with decreasing $V_{\rm rot}$ rather than simply an abrupt change at a given galactic mass.

It can also be seen from Tables \ref{fit} and \ref{prop} that in order to model the FIR emission of UGC 7321 and NGC 4244, which are known to contain molecular hydrogen in the nuclear regions of their disks, it has been necessary to include a larger fraction of emission from our template of embedded compact star forming regions. IC 2233 on the other hand has not been detected in CO emission and, indeed, is shown to require a much lower amount of emission to originate from compact sources with a warmer dust emission peak. This result is not surprising when one examines high angular resolution optical imaging of UGC 7321 and NGC 4244 (Figure \ref{HST}) in which dark clouds suggestive of molecular gas are observed. IC 2233 shows some dark clouds in its central regions but less than the other two galaxies studied, reinforcing the suggestion that it is the galaxy poorest in interstellar molecular gas.

From our total dust masses (Table \ref{fit}) and the \HI{} and molecular hydrogen masses from the literature (Table \ref{prop}) we can calculate global gas-to-dust ratios\footnote{In the case of IC 2233 we use the upper limit $H_{2}$ mass of $1.4\times10^{6}$M$_{\odot}$}. After accounting for the contribution of Helium (assuming $Y=0.25$) we find global gas-to-dust ratios of $\sim 666$ (UGC 7321), $734$ (NGC 4244) and $1266$ (IC 2233). These values are significantly higher than the Galactic value of $\sim 140$ \citep{whittet_dust_1992}. The gas-to-dust ratios of UGC 7321 and NGC 4244 fall between those found for the Large Magellanic Cloud \citep{koornneef_gas_1982} and the Small Magellanic Cloud (SMC) \citep{bouchet_visible_1985} while the value for IC 2233 is consistent with that of the SMC.

\subsection{Model Limitations and Future Prospects}
It is somewhat unclear what effect the adoption of a clumpy dust and emissivity distribution would have on the FIR emission of our galaxy sample. \citet{matthews_modeling_2001} found that in order to reproduce the optical appearance of UGC 7321 it was necessary to allocate approximately $50\%$ of the dust mass to a clumpy component. The effect that such a two phase diffuse dust distribution may have on the FIR emission would depend on the associated emissivity distributions. Such dense clumps could be ``quiescent'' and represent over-dense regions of the ISM illuminated only by the diffuse ISRF, but recent evidence suggests that such clouds are optically thin and will contribute little to the attenuation and dust emission \citep{molinari_clouds_2010}. However, if the dark clouds seen in $HST$ imaging are star forming molecular clouds then they will be illuminated by both the ISRF and by young stars embedded within them. It is thought that in the case of LSB galaxies the optical depths of the clumps are similar to those found for diffuse molecular clouds in the Milky Way \citep{matthews_modeling_2001}. In this case the clumpy component may provide a significant contribution to the FIR emission reducing the need to have a radially extended dust disk as is found using smooth density distributions.

It has also been shown by \citet{indebetouw_three-dimensional_2006} that the observed SED of young stellar objects (YSOs) that are embedded in clumpy circumstellar material can have a strong dependence on the observer's viewing angle (azimuth) and inclination. The observed MIR flux was found to vary by up to two orders of magnitude for the same highly embedded system viewed along a different sightline. The effects are more severe at shorter wavelengths. Although the case of YSOs is very different to the galactic environment studied here it should be noted that the effects of a clumpy galactic ISM containing optically thick clouds may cause viewing angle and inclination dependent effects on the MIR SEDs that are not accounted for in our smooth, axisymmetric models.

Currently we have no observations of our galaxy sample in the sub-mm wavelength regime. It has been found that in order to reproduce the sub-mm emission observed from some HSB galaxies it becomes necessary to include an additional dust mass \citep{popescu_modelling_2000,misiriotis_modeling_2001,popescu_modelling_2011}. The additional dust mass reveals its presence through sub-mm emission that is underestimated by models which can adequately describe the SED to $\sim 100\mu$m. Further evidence for an extended cold dust component is also found by \citet{holwerda_opacity_2005} based on attenuation of background galaxies in face-on disks. Until observations become available we cannot be sure that our models do not lack cold dust emission at longer wavelengths.

\section{Summary}
We have utilized multi-wavelength imaging and photometry in conjunction with sophisticated Monte Carlo radiation transfer codes to investigate the structure of three edge-on, LSB disk galaxies. The galaxies have been chosen to span a range in central, optical surface brightness and molecular hydrogen masses. 

We have been able to reproduce the global, optical appearance of all three galaxies using smooth emissivity and dust distributions. We find that the composition and size distribution of dust grains adopted, which are based on Milky Way extinction, provide a good match to the observed properties in our sample of LSB disk galaxies. Our models also reproduce the total emission at $70$ and $160\mu$m for all three galaxies. However, the FIR morphology of our models appears more centrally concentrated than the more diffuse distribution suggested by the data. We find that the dust mass appears to be distributed in an exponential disk with a scale height comparable to or exceeding that of the stellar disk. This is in contrast to the findings for HSB galaxies where the dust disk is found have a vertical scale height of approximately half the stellar disk \citep{xilouris_are_1999}. The comparable scale heights in the dust and stellar disks is likely associated with the increased stability of the ISM in LSB disks against vertical collapse \citep{dalcanton_formation_2004} and the thin nature of the stellar disks, which suggests minimal dynamical heating.

The dust masses and distributions derived suggest dust masses in the range $1.16-2.38\times 10^{6} M_{\odot}$ corresponding to face on, $V$ band, optical depths between $\tau_{face}=0.034-0.106$. 

In future work we hope to develop our radiation transfer models to include small scale non-axisymmetric structures which may shed further light on the structure of the ISM and star formation processes in LSB disk galaxies. The inclusion of radial variations in the dust disk scale heights, associated with a flaring gas disk, could also prove important. Additional sub-mm observations may also allow us to uncover cold dust that is associated with the extended \HI{}, as has been found in some HSB galaxies \citep{popescu_first_2003,hinz_extended_2006}.

\begin{acknowledgements}
We would like to thank the anonymous referee for helpful comments that improved many aspects of this work.

J.M.M. wishes to thank Aaron Robotham and Lee Kelvin for useful discussions on many aspects of this work as well as Cristina Popescu for helpful comments.

Financial support for this work was provided to L. D. M. through contract 1279242 from the Jet Propulsion Laboratory (JPL). This work is based in part on observations from the {\it Spitzer Space Telescope}, operated by JPL, California Institute of Technology, under contract 1407 with the National Aeronautic and Space Administration (NASA)

J.S.G. thanks the University of Wisconsin Graduate School for partial support of this research.

Funding for the SDSS and SDSS-II has been provided by the Alfred P. Sloan Foundation, the Participating Institutions, the National Science Foundation, the U.S. Department of Energy, the National Aeronautics and Space Administration, the Japanese Monbukagakusho, the Max Planck Society, and the Higher Education Funding Council for England. The SDSS Web Site is http://www.sdss.org/.

The SDSS is managed by the Astrophysical Research Consortium for the Participating Institutions. The Participating Institutions are the American Museum of Natural History, Astrophysical Institute Potsdam, University of Basel, University of Cambridge, Case Western Reserve University, University of Chicago, Drexel University, Fermilab, the Institute for Advanced Study, the Japan Participation Group, Johns Hopkins University, the Joint Institute for Nuclear Astrophysics, the Kavli Institute for Particle Astrophysics and Cosmology, the Korean Scientist Group, the Chinese Academy of Sciences (LAMOST), Los Alamos National Laboratory, the Max-Planck-Institute for Astronomy (MPIA), the Max-Planck-Institute for Astrophysics (MPA), New Mexico State University, Ohio State University, University of Pittsburgh, University of Portsmouth, Princeton University, the United States Naval Observatory, and the University of Washington.

This work makes use of observations made with the NASA/ESA {\it Hubble Space Telescope}, and obtained from the Hubble Legacy Archive, which is a collaboration between the Space Telescope Science Institute (STScI/NASA), the Space Telescope European Coordinating Facility (ST-ECF/ESA) and the Canadian Astronomy Data Centre (CADC/NRC/CSA). 

{\it GALEX} is a NASA Small Explorer, and we gratefully acknowledge NASAÕs support for construction, operation, and science analysis for the {\it GALEX} mission. 

This publication makes use of data products from the Two Micron All Sky Survey, which is a joint project of the University of Massachusetts and the Infrared Processing and Analysis Center/California Institute of Technology, funded by the NASA and the National Science Foundation. 

This research made use of the NASA / IPAC Extragalactic Database, which is operated by JPL/Caltech, under contract with NASA.
\end{acknowledgements}
\bibliographystyle{apj}
\bibliography{Mylibrary}

\begin{thebibliography}{77}
\expandafter\ifx\csname natexlab\endcsname\relax\def\natexlab#1{#1}\fi

\bibitem[{Abazajian {et~al.}(2009)Abazajian, {Adelman-McCarthy}, Agüeros,
  Allam, Allende~Prieto, An, Anderson, Anderson, Annis, Bahcall,
  {Bailer-Jones}, Barentine, Bassett, Becker, Beers, Bell, Belokurov, Berlind,
  Berman, Bernardi, Bickerton, Bizyaev, Blakeslee, Blanton, Bochanski, Boroski,
  Brewington, Brinchmann, Brinkmann, Brunner, Budavári, Carey, Carliles, Carr,
  Castander, Cinabro, Connolly, Csabai, Cunha, Czarapata, Davenport, de~Haas,
  Dilday, Doi, Eisenstein, Evans, Evans, Fan, Friedman, Frieman, Fukugita,
  Gänsicke, Gates, Gillespie, Gilmore, Gonzalez, Gonzalez, Grebel, Gunn,
  Györy, Hall, Harding, Harris, Harvanek, Hawley, Hayes, Heckman, Hendry,
  Hennessy, Hindsley, Hoblitt, Hogan, Hogg, Holtzman, Hyde, Ichikawa, Ichikawa,
  Im, Ivezić, Jester, Jiang, Johnson, Jorgensen, Jurić, Kent, Kessler,
  Kleinman, Knapp, Konishi, Kron, Krzesinski, Kuropatkin, Lampeitl, Lebedeva,
  Lee, Lee, Leger, Lépine, Li, Lima, Lin, Long, Loomis, Loveday, Lupton,
  Magnier, Malanushenko, Malanushenko, Mandelbaum, Margon, Marriner,
  {Martínez-Delgado}, Matsubara, {McGehee}, {McKay}, Meiksin, Morrison,
  Mullally, Munn, Murphy, Nash, Nebot, Neilsen, Newberg, Newman, Nichol,
  Nicinski, {Nieto-Santisteban}, Nitta, Okamura, Oravetz, Ostriker, Owen,
  Padmanabhan, Pan, Park, Pauls, Peoples, Percival, Pier, Pope, Pourbaix,
  Price, Purger, Quinn, Raddick, Fiorentin, Richards, Richmond, Riess, Rix,
  Rockosi, Sako, Schlegel, Schneider, Scholz, Schreiber, Schwope, Seljak,
  Sesar, Sheldon, Shimasaku, Sibley, Simmons, Sivarani, Smith, Smith,
  Smolčić, Snedden, Stebbins, Steinmetz, Stoughton, Strauss, Subba~Rao, Suto,
  Szalay, Szapudi, Szkody, Tanaka, Tegmark, Teodoro, Thakar, Tremonti, Tucker,
  Uomoto, Vanden~Berk, Vandenberg, Vidrih, Vogeley, Voges, Vogt, Wadadekar,
  Watters, Weinberg, West, White, Wilhite, Wonders, Yanny, Yocum, York, Zehavi,
  Zibetti, \& Zucker}]{abazajian_seventh_2009}
Abazajian, K.~N., Adelman-McCarthy, J.~K. and Ag\"ueros, M.~A., {et~al.} 2009, \apjs, 182, 543

\bibitem[{Auld {et~al.}(2006)Auld, de~Blok, Bell, \&
  Davies}]{auld_morphology_2006}
Auld, R., de~Blok, W. J.~G., Bell, E., \& Davies, J.~I. 2006, \mnras, 366, 1475

\bibitem[{Baes {et~al.}(2010)Baes, Fritz, Gadotti, Smith, Dunne, da~Cunha,
  Amblard, Auld, Bendo, Bonfield, Burgarella, Buttiglione, Cava, Clements,
  Cooray, Dariush, de~Zotti, Dye, Eales, Frayer, {Gonzalez-Nuevo}, Herranz,
  Ibar, Ivison, Lagache, Leeuw, {Lopez-Caniego}, Jarvis, Maddox, Negrello,
  Michałowski, Pascale, Pohlen, Rigby, Rodighiero, Samui, Serjeant, Temi,
  Thompson, van~der Werf, Verma, \& Vlahakis}]{baes_herschel-atlas:_2010}
Baes, M., Fritz, J., Gadotti, D.~A., {et~al.} 2010, \aap, 518, L39

\bibitem[{Banerjee {et~al.}(2010)Banerjee, Matthews, \&
  Jog}]{banerjee_dark_2010}
Banerjee, A., Matthews, L.~D., \& Jog, C.~J. 2010, NewA, 15, 89

\bibitem[{Bianchi(2007)}]{bianchi_dust_2007}
Bianchi, S. 2007, \aap, 471, 765

\bibitem[{Bianchi(2008)}]{bianchi_dust_2008}
---. 2008, \aap, 490, 461

\bibitem[{Bianchi {et~al.}(2000)Bianchi, Ferrara, Davies, \&
  Alton}]{bianchi_effects_2000}
Bianchi, S., Ferrara, A., Davies, J.~I., \& Alton, P.~B. 2000, \mnras, 311, 601

\bibitem[{Bizyaev \& Kajsin(2004)}]{bizyaev_stellar_2004}
Bizyaev, D., \& Kajsin, S. 2004, \apj, 613, 886

\bibitem[{Bjorkman \& Wood(2001)}]{bjorkman_radiative_2001}
Bjorkman, J.~E., \& Wood, K. 2001, \apj, 554, 615

\bibitem[{Bouchet {et~al.}(1985)Bouchet, Lequeux, Maurice, Prevot, \&
  {Prevot-Burnichon}}]{bouchet_visible_1985}
Bouchet, P., Lequeux, J., Maurice, E., Prevot, L., \& {Prevot-Burnichon}, M.~L.
  1985, \aap, 149, 330

\bibitem[{Cayatte {et~al.}(1994)Cayatte, Kotanyi, Balkowski, \& van
  Gorkom}]{cayatte_very_1994}
Cayatte, V., Kotanyi, C., Balkowski, C., \& van Gorkom, J.~H. 1994, \aj, 107,
  1003

\bibitem[{Chini {et~al.}(1986)Chini, Kreysa, Mezger, \&
  Gemuend}]{chini_1.3_1986}
Chini, R., Kreysa, E., Mezger, P.~G., \& Gemuend, H. 1986, \aap, 154, L8

\bibitem[{Dalcanton {et~al.}(2004)Dalcanton, Yoachim, \&
  Bernstein}]{dalcanton_formation_2004}
Dalcanton, J.~J., Yoachim, P., \& Bernstein, R.~A. 2004, \apj, 608, 189

\bibitem[{Dale {et~al.}(2009)Dale, Cohen, Johnson, Schuster, Calzetti,
  Engelbracht, Gil~de Paz, Kennicutt, Lee, Begum, Block, Dalcanton, Funes,
  Gordon, Johnson, Marble, Sakai, Skillman, van Zee, Walter, Weisz, Williams,
  Wu, \& Wu}]{dale_Spitzer_2009}
Dale, D.~A., Cohen, S.~A., Johnson, L.~C., {et~al.} 2009, \apj, 703, 517

\bibitem[{Das {et~al.}(2006)Das, {O'Neil}, Vogel, \& {McGaugh}}]{das_CO_2006}
Das, M., {O'Neil}, K., Vogel, S.~N., \& {McGaugh}, S. 2006, \apj, 651, 853

\bibitem[{de~Blok \& Bosma(2002)}]{de_blok_high-resolution_2002}
de~Blok, W. J.~G., \& Bosma, A. 2002, \aap, 385, 816

\bibitem[{de~Blok \& {McGaugh}(1997)}]{de_blok_dark_1997}
de~Blok, W. J.~G., \& {McGaugh}, S.~S. 1997, \mnras, 290, 533

\bibitem[{Draine \& Li(2007)}]{draine_infrared_2007}
Draine, B.~T., \& Li, A. 2007, \apj, 657, 810

\bibitem[{Driver(1999)}]{driver_contribution_1999}
Driver, S.~P. 1999, \apj, 526, L69

\bibitem[{Engelbracht {et~al.}(2005)Engelbracht, Gordon, Rieke, Werner, Dale,
  \& Latter}]{engelbracht_metallicity_2005}
Engelbracht, C.~W., Gordon, K.~D., Rieke, G.~H., {et~al.} 2005, \apj, 628, L29

\bibitem[{Fry {et~al.}(1999)Fry, Morrison, Harding, \& Boroson}]{fry_deep_1999}
Fry, A.~M., Morrison, H.~L., Harding, P., \& Boroson, T.~A. 1999, \aj, 118,
  1209

\bibitem[{Galliano {et~al.}(2008)Galliano, Dwek, \&
  Chanial}]{galliano_stellar_2008}
Galliano, F., Dwek, E., \& Chanial, P. 2008, \apj, 672, 214

\bibitem[{Gordon {et~al.}(2008)Gordon, Engelbracht, Rieke, Misselt, Smith, \&
  Kennicutt}]{gordon_behavior_2008}
Gordon, K.~D., Engelbracht, C.~W., Rieke, G.~H., {et~al.} 2008, \apj, 682, 336

\bibitem[{Hinz {et~al.}(2006)Hinz, Misselt, Rieke, Rieke, Smith, Blaylock, \&
  Gordon}]{hinz_extended_2006}
Hinz, J.~L., Misselt, K., Rieke, M.~J., {et~al.} 2006, \apj, 651, 874

\bibitem[{Holwerda {et~al.}(2005)Holwerda, González, van~der Kruit, \&
  Allen}]{holwerda_opacity_2005}
Holwerda, B.~W., González, R.~A., van~der Kruit, P.~C., \& Allen, R.~J. 2005,
  \aap, 444, 109

\bibitem[{Hoopes {et~al.}(1999)Hoopes, Walterbos, \&
  Rand}]{hoopes_diffuse_1999}
Hoopes, C.~G., Walterbos, R. A.~M., \& Rand, R.~J. 1999, \apj, 522, 669

\bibitem[{Indebetouw {et~al.}(2006)Indebetouw, Whitney, Johnson, \&
  Wood}]{indebetouw_three-dimensional_2006}
Indebetouw, R., Whitney, B.~A., Johnson, K.~E., \& Wood, K. 2006, \apj, 636,
  362

\bibitem[{Jimenez {et~al.}(1998)Jimenez, Padoan, Matteucci, \&
  Heavens}]{jimenez_galaxy_1998}
Jimenez, R., Padoan, P., Matteucci, F., \& Heavens, A.~F. 1998, \mnras, 299,
  123

\bibitem[{Jonsson {et~al.}(2010)Jonsson, Groves, \&
  Cox}]{jonsson_high-resolution_2010}
Jonsson, P., Groves, B.~A., \& Cox, T.~J. 2010, \mnras, 403, 17

\bibitem[{Kennicutt(1989)}]{kennicutt_star_1989}
Kennicutt, R.~C. 1989, \apj, 344, 685

\bibitem[{Kennicutt(1998)}]{kennicutt_star_1998}
---. 1998, \araa, 36, 189

\bibitem[{Kennicutt {et~al.}(2008)Kennicutt, Lee, Funes, Sakai, \&
  Akiyama}]{kennicutt_h_2008}
Kennicutt, R.~C., Lee, J.~C., Funes, S.~J., Sakai, S., \& Akiyama, S. 2008,
  \apjs, 178, 247

\bibitem[{Kim {et~al.}(1994)Kim, Martin, \& Hendry}]{kim_size_1994}
Kim, S., Martin, P.~G., \& Hendry, P.~D. 1994, \apj, 422, 164

\bibitem[{Kodaira \& Yamashita(1996)}]{kodaira_near-infrared_1996}
Kodaira, K., \& Yamashita, T. 1996, \pasj, 48, 581

\bibitem[{Koornneef(1982)}]{koornneef_gas_1982}
Koornneef, J. 1982, \aap, 107, 247

\bibitem[{Kotulla {et~al.}(2009)Kotulla, Fritze, Weilbacher, Anders, \&
  the~galev team}]{kotulla_GALEV_2009}
Kotulla, R., Fritze, U., Weilbacher, P., Anders, P., \& the~galev team. 2009,
  \mnras, 396, 462

\bibitem[{Kurucz(1993)}]{kurucz_kurucz_1993}
Kurucz, R. 1993, {KURUCZ} {CD-ROM} No. 13, {ATLAS9} Stellar Atmosphere Programs
  and 2 km/s grid {(Cambridge:SAO)}

\bibitem[{Kylafis \& Bahcall(1987)}]{kylafis_dust_1987}
Kylafis, N.~D., \& Bahcall, J.~N. 1987, \apj, 317, 637

\bibitem[{Lee {et~al.}(2011)Lee, Gil~de Paz, Kennicutt, Bothwell, Dalcanton,
  José G. Funes~S., Johnson, Sakai, Skillman, Tremonti, \& van
  Zee}]{lee_galex_2011}
Lee, J.~C., Gil de Paz, A., Kennicutt, Jr., R.~C., {et~al.} 2011, \apjs, 192, 6

\bibitem[{Madden {et~al.}(2006)Madden, Galliano, Jones, \&
  Sauvage}]{madden_ism_2006}
Madden, S.~C., Galliano, F., Jones, A.~P., \& Sauvage, M. 2006, \aap, 446, 877

\bibitem[{Mathis {et~al.}(1983)Mathis, Mezger, \&
  Panagia}]{mathis_interstellar_1983}
Mathis, J.~S., Mezger, P.~G., \& Panagia, N. 1983, \aap, 128, 212

\bibitem[{Matthews {et~al.}(1999)Matthews, Gallagher, \& van
  Driel}]{matthews_extraordinary_1999}
Matthews, L.~D., Gallagher, J.~S., \& van Driel, W. 1999, \aj, 118, 2751

\bibitem[{Matthews \& Gao(2001)}]{matthews_co_2001}
Matthews, L.~D., \& Gao, Y. 2001, \apj, 549, L191

\bibitem[{Matthews {et~al.}(2005)Matthews, Gao, Uson, \&
  Combes}]{matthews_detections_2005}
Matthews, L.~D., Gao, Y., Uson, J.~M., \& Combes, F. 2005, \aj, 129, 1849

\bibitem[{Matthews \& Uson(2008)}]{matthews_H_2008}
Matthews, L.~D., \& Uson, J.~M. 2008, \aj, 135, 291

\bibitem[{Matthews \& Wood(2001)}]{matthews_modeling_2001}
Matthews, L.~D., \& Wood, K. 2001, \apj, 548, 150

\bibitem[{Matthews \& Wood(2003)}]{matthews_high-latitude_2003}
---. 2003, \apj, 593, 721

\bibitem[{{McGaugh}(1994)}]{mcgaugh_oxygen_1994}
{McGaugh}, S.~S. 1994, \apj, 426, 135

\bibitem[{{McGaugh} {et~al.}(1995){McGaugh}, Schombert, \&
  Bothun}]{mcgaugh_morphology_1995}
{McGaugh}, S.~S., Schombert, J.~M., \& Bothun, G.~D. 1995, \aj, 109, 2019

\bibitem[{Minchin {et~al.}(2004)Minchin, Disney, Parker, Boyce, de~Blok, Banks,
  Ekers, Freeman, Garcia, Gibson, Grossi, Haynes, Knezek, Lang, Malin, Price,
  Putman, Stewart, \& Wright}]{minchin_cosmological_2004}
Minchin, R.~F., Disney, M.~J., Parker, Q.~A., {et~al.} 2004, \mnras, 355, 1303

\bibitem[{Misiriotis {et~al.}(2001)Misiriotis, Popescu, Tuffs, \&
  Kylafis}]{misiriotis_modeling_2001}
Misiriotis, A., Popescu, C.~C., Tuffs, R., \& Kylafis, N.~D. 2001, \aap, 372,
  775

\bibitem[{Molinari {et~al.}(2010)Molinari, Swinyard, Bally, Barlow, Bernard,
  Martin, Moore, {Noriega-Crespo}, Plume, Testi, Zavagno, Abergel, Ali,
  Anderson, André, Baluteau, Battersby, Beltrán, Benedettini, Billot,
  Blommaert, Bontemps, Boulanger, Brand, Brunt, Burton, Calzoletti, Carey,
  Caselli, Cesaroni, Cernicharo, Chakrabarti, Chrysostomou, Cohen, Compiegne,
  de~Bernardis, de~Gasperis, di~Giorgio, Elia, Faustini, Flagey, Fukui, Fuller,
  Ganga, {Garcia-Lario}, Glenn, Goldsmith, Griffin, Hoare, Huang, Ikhenaode,
  Joblin, Joncas, Juvela, Kirk, Lagache, Li, Lim, Lord, Marengo, Marshall,
  Masi, Massi, Matsuura, Minier, {Miville-Deschênes}, Montier, Morgan, Motte,
  Mottram, Müller, Natoli, Neves, Olmi, Paladini, Paradis, Parsons, Peretto,
  Pestalozzi, Pezzuto, Piacentini, Piazzo, Polychroni, Pomarès, Popescu,
  Reach, Ristorcelli, Robitaille, Robitaille, Rodón, Roy, Royer, Russeil,
  Saraceno, Sauvage, Schilke, Schisano, Schneider, Schuller, Schulz, Sibthorpe,
  Smith, Smith, Spinoglio, Stamatellos, Strafella, Stringfellow, Sturm, Taylor,
  Thompson, Traficante, Tuffs, Umana, Valenziano, Vavrek, Veneziani, Viti,
  Waelkens, {Ward-Thompson}, White, Wilcock, Wyrowski, Yorke, \&
  Zhang}]{molinari_clouds_2010}
Molinari, S., Swinyard, B., Bally, J., {et~al.} 2010, \aap, 518, L100

\bibitem[{Morrissey {et~al.}(2005)Morrissey, Schiminovich, Barlow, Martin,
  Blakkolb, Conrow, Cooke, Erickson, Fanson, Friedman, Grange, Jelinsky, Lee,
  Liu, Mazer, {McLean}, Milliard, Randall, Schmitigal, Sen, Siegmund, Surber,
  Vaughan, Viton, Welsh, Bianchi, Byun, Donas, Forster, Heckman, Lee, Madore,
  Malina, Neff, Rich, Small, Szalay, \& Wyder}]{morrissey_-orbit_2005}
Morrissey, P., Schiminovich, D., Barlow, T.~A., {et~al.} 2005, \apj, 619, L7

\bibitem[{{Muñoz-Mateos} {et~al.}(2009){Muñoz-Mateos}, Gil~de Paz, Boissier,
  Zamorano, Dale, {Pérez-González}, Gallego, Madore, Bendo, Thornley, Draine,
  Boselli, Buat, Calzetti, Moustakas, \& Kennicutt}]{munoz-mateos_radial_2009}
Mu\~noz-Mateos, J.~C., Gil de Paz, A., Boissier, S., {et~al.} 2009, \apj, 701, 1965

\bibitem[{Olling(1996)}]{olling_ngc_1996}
Olling, R.~P. 1996, \aj, 112, 457

\bibitem[{{O'Neil} \& Bothun(2000)}]{oneil_space_2000}
{O'Neil}, K., \& Bothun, G. 2000, \apj, 529, 811

\bibitem[{Pierini {et~al.}(2004)Pierini, Gordon, Witt, \&
  Madsen}]{pierini_dust_2004}
Pierini, D., Gordon, K.~D., Witt, A.~N., \& Madsen, G.~J. 2004, \apj, 617, 1022

\bibitem[{Popescu {et~al.}(2000)Popescu, Misiriotis, Kylafis, Tuffs, \&
  Fischera}]{popescu_modelling_2000}
Popescu, C.~C., Misiriotis, A., Kylafis, N.~D., Tuffs, R.~J., \& Fischera, J.
  2000, \aap, 362, 138

\bibitem[{Popescu \& Tuffs(2003)}]{popescu_first_2003}
Popescu, C.~C., \& Tuffs, R.~J. 2003, \aap, 410, L21

\bibitem[{Popescu {et~al.}(2011)Popescu, Tuffs, Dopita, Fischera, Kylafis, \&
  Madore}]{popescu_modelling_2011}
Popescu, C.~C., Tuffs, R.~J., Dopita, M.~A., {et~al.} 2011, \aap, 527, 109

\bibitem[{Rand(1996)}]{rand_diffuse_1996}
Rand, R.~J. 1996, \apj, 462, 712

\bibitem[{Sage(1993)}]{sage_molecular_1993}
Sage, L.~J. 1993, \aap, 272, 123

\bibitem[{Savoy {et~al.}(2009)Savoy, Welch, \& Fich}]{savoy_scuba_2009}
Savoy, J., Welch, G.~A., \& Fich, M. 2009, \apj, 706, 21

\bibitem[{Seth {et~al.}(2005{\natexlab{a}})Seth, Dalcanton, \&
  de~Jong}]{seth_study_2005-1}
Seth, A.~C., Dalcanton, J.~J., \& de~Jong, R.~S. 2005{\natexlab{a}}, \aj, 129,
  1331

\bibitem[{Seth {et~al.}(2005{\natexlab{b}})Seth, Dalcanton, \&
  de~Jong}]{seth_study_2005}
---. 2005{\natexlab{b}}, \aj, 130, 1574

\bibitem[{Sprayberry {et~al.}(1997)Sprayberry, Impey, Irwin, \&
  Bothun}]{sprayberry_low_1997}
Sprayberry, D., Impey, C.~D., Irwin, M.~J., \& Bothun, G.~D. 1997, \apj, 482,
  104

\bibitem[{Tempel {et~al.}(2010)Tempel, Tamm, \&
  Tenjes}]{tempel_dust-corrected_2010}
Tempel, E., Tamm, A., \& Tenjes, P. 2010, \aap, 509, 91

\bibitem[{Uson \& Matthews(2003)}]{uson_H_2003}
Uson, J.~M., \& Matthews, L.~D. 2003, \aj, 125, 2455

\bibitem[{van~der Hulst {et~al.}(1993)van~der Hulst, Skillman, Smith, Bothun,
  {McGaugh}, \& de~Blok}]{van_der_hulst_star_1993}
van~der Hulst, J.~M., Skillman, E.~D., Smith, T.~R., {et~al.} 1993, \aj, 106, 548

\bibitem[{Vorobyov {et~al.}(2009)Vorobyov, Shchekinov, Bizyaev, Bomans, \&
  Dettmar}]{vorobyov_age_2009}
Vorobyov, E.~I., Shchekinov, Y., Bizyaev, D., Bomans, D., \& Dettmar, R. 2009,
  \aap, 505, 483

\bibitem[{Whittet(1992)}]{whittet_dust_1992}
Whittet, D. C.~B. 1992, Dust in the galactic environment ({Bristol:IOP})

\bibitem[{Wood {et~al.}(2008)Wood, Whitney, Robitaille, \&
  Draine}]{wood_emission_2008}
Wood, K., Whitney, B.~A., Robitaille, T., \& Draine, B.~T. 2008, \apj, 688,
  1118

\bibitem[{Xilouris {et~al.}(1998)Xilouris, Alton, Davies, Kylafis,
  Papamastorakis, \& Trewhella}]{xilouris_optical_1998}
Xilouris, E.~M., Alton, P.~B., Davies, J.~I., {et~al.} 1998, \aap, 331, 894–900

\bibitem[{Xilouris {et~al.}(1999)Xilouris, Byun, Kylafis, Paleologou, \&
  Papamastorakis}]{xilouris_are_1999}
Xilouris, E.~M., Byun, Y.~I., Kylafis, N.~D., Paleologou, E.~V., \&
  Papamastorakis, J. 1999, \aap, 344, 868

\bibitem[{Xilouris {et~al.}(1997)Xilouris, Kylafis, Papamastorakis, Paleologou,
  \& Haerendel}]{xilouris_distribution_1997}
Xilouris, E.~M., Kylafis, N.~D., Papamastorakis, J., Paleologou, E.~V., \&
  Haerendel, G. 1997, \aap, 325, 135–143

\bibitem[{Young \& Scoville(1991)}]{young_molecular_1991}
Young, J.~S., \& Scoville, N.~Z. 1991, \araa, 29, 581

\bibitem[{Zackrisson {et~al.}(2005)Zackrisson, Bergvall, \&
  \"Ostlin}]{zackrisson_stellar_2005}
Zackrisson, E., Bergvall, N., \& \"Ostlin, G. 2005, \aap, 435, 29

\end{thebibliography}

\end{document}